\documentclass[showpacs,preprintnumbers,amsmath,amssymb,preprint]{revtex4}

\pdfoutput=1

   \usepackage{graphicx}
\usepackage{latexsym}
\usepackage{natbib}

\usepackage{amssymb}

\usepackage{array}
\newcolumntype{L}[1]{>{\raggedright\let\newline\\\arraybackslash\hspace{0pt}}m{#1}}
\newcolumntype{C}[1]{>{\centering\let\newline\\\arraybackslash\hspace{0pt}}m{#1}}
\newcolumntype{R}[1]{>{\raggedleft\let\newline\\\arraybackslash\hspace{0pt}}m{#1}}

\usepackage{multirow}
\usepackage{dcolumn}
\usepackage{tabularx}

\usepackage{amssymb,amsmath}
\usepackage{amsthm}

\usepackage{xcolor}

\usepackage{alltt} 

\usepackage{float}
\floatstyle{ruled}
\newfloat{Input File}{thp}{loi}

\usepackage{url}

\newcommand{\Dpx}{D_{+,v_x}}
\newcommand{\Dmx}{D_{-,v_x}}
\newcommand{\Dpy}{D_{+,v_y}}
\newcommand{\Dmy}{D_{-,v_y}}
\newcommand{\Dpmx}{D_{\pm,v_x}}

\newcommand{\SDpx}{{\mathcal{D}}_{+,v_x}}
\newcommand{\SDmx}{{\mathcal{D}}_{-,v_x}}
\newcommand{\SDpmx}{{\mathcal{D}}_{\pm,v_x}}

\newcommand{\Dzx}{D_{0,v_x}}
\newcommand{\Dzy}{D_{0,v_y}}
\newcommand{\SDzx}{{\mathcal{D}}_{0,v_x}}
\newcommand{\SDzy}{{\mathcal{D}}_{0,v_y}}

\newcommand{\Order}{{\mathcal{O}}}

\newcommand{\dvx}{\Delta v_x}
\newcommand{\dvy}{\Delta v_y}

\newcommand{\pd}{\partial}
\newcommand{\pt}[1]{\frac{\pd #1}{\pd t}}
\newcommand{\px}[1]{\frac{\pd #1}{\pd x}}
\newcommand{\py}[1]{\frac{\pd #1}{\pd y}}
\newcommand{\pvx}[1]{\frac{\pd #1}{\pd v_x}}
\newcommand{\pvy}[1]{\frac{\pd #1}{\pd v_y}}

\def \x {\vec{x}}
\def \v {\vec{v}}

\def \vth {v_{\rm th}}
\def \vthe {v_{\rm th\,e}}
\def \vthi {v_{\rm th\,i}}

\def \nuth {\nu_{ei,th}}
\def \vph {v_{\phi}}

\draft 

\begin{document}

\preprint{LLNL-JRNL-669228}

\title{Vlasov Simulations of Electron-Ion Collision Effects on Damping of
  Electron Plasma Waves} \thanks{This work was performed under the auspices of
  the U.S. Department of Energy by Lawrence Livermore National Laboratory
  under contract number DE-AC52-07NA27344.  This work was funded by the
  Laboratory Directed Research and Development Program at LLNL under project
  tracking codes 12-ERD-061 and 15-ERD-038.}

\author{J. W. Banks$^1$}
\email[]{banksj3@rpi.edu}

\author{S. Brunner$^2$}

\author{R. L. Berger$^3$}

\author{T. M. Tran$^2$}
\affiliation{
  (1) Rensselaer Polytechnic Institute,  \\
  Department of Mathematical Sciences, Troy, NY 12180}
\affiliation{  
  (2) Ecole Polytechnique F\'ed\'erale de Lausanne (EPFL),\\
Swiss Plasma Center (SPC),\\
CH-1015 Lausanne, Switzerland}
\affiliation{
  (3) Lawrence Livermore National Laboratory,
  P.O. Box 808,
  Livermore, California 94551
}

\date{\today}

\begin{abstract}
Collisional effects can play an essential role in the dynamics of plasma waves
by setting a minimum damping rate and by interfering with wave-particle
resonances.  Kinetic simulations of the effects of electron-ion pitch angle
scattering on Electron Plasma Waves (EPWs) are presented here. In particular,
the effects of such collisions on the frequency and damping of small-amplitude
EPWs for a range of collision rates and wave phase velocities are computed and
compared with theory. 
Both the Vlasov simulations and linear kinetic theory find the direct contribution of electron-ion 
collisions to wave damping is about a factor of two smaller than is obtained from linearized 
fluid theory. To our knowledge, this simple result has not been published before.

Simulations have been carried out using
a grid-based (Vlasov) approach, based on a high-order conservative finite
difference method for discretizing the Fokker-Planck equation describing the
evolution of the electron distribution function. Details of the implementation
of the collision operator within this framework are presented. Such a
grid-based approach, which is not subject to numerical noise, is of particular
interest for the accurate measurements of the wave damping rates.
\end{abstract}


\maketitle

\section{Introduction}

Simulations of kinetic processes in plasmas make use of either
Particle-in-Cell (PIC) methods or direct discretization of the Vlasov equation
based on an Eulerian (or grid-based) representation. Computations carried out
with the latter approach are also referred to as {\it Vlasov} simulations. In
the context of PIC simulations of laser plasma interactions, methods for
including collisional effects have been implemented and
validated~\cite{Abe1977, Cohen10, Dimits13}, and the effect of their inclusion
has been shown in studies of stimulated Brillouin scattering with ion-ion
collisions,\cite{Cohen06} simulations of fast ignition with electron
collisions,\cite{Kemp2010} and with electron-ion, electron-electron and
ion-ion collisions in simulations of counter-streaming plasma
flows.\cite{Ross2013} Recent developments have seen the application of
multidimensional grid-based simulation,\cite{Banks2011,Winjum2013,
  Berger2015} but collisions were not included in these studies.

When collisional effects are strong enough to enforce a nearly
isotropic or Maxwell-Boltzmann distribution, methods such as 
'Fokker-Planck',\cite{Epperlein94, Brunner02, Tzoufras2011}
hydrodynamic descriptions,\cite{Braginskii1965} or nonlocal hydrodynamic
descriptions~\cite{Albritton1986, Epperlein1991, Bychenkov1995, Brantov1996,
  Schurtz2000} may be applicable. Collisional effects may be
important to consider in kinetic simulations 
as they can set a minimum damping rate for high phase velocity Electron Plasma (or Langmuir) Waves
(EPWs) and ion acoustic waves in single-ion species \cite{Brantov12,
  Epperlein1992} or multiple-ion species plasmas.\cite{Berger2005} In inertial
fusion applications, collisional damping $\nu_{\rm coll}$ of EPWs is very
weak, \textit{e.g.}, $\nu_{\rm coll}/\omega_{pe} \simeq \nu_{ei}^{\rm brag}
/(2\omega_{pe}) \sim 10^{-4}$ for a helium plasma at electron temperature $T_e
= 2.5~keV$ and electron density $N_e = Z N_i = 9\cdot10^{20}~cm^{-3}$ where
$Z$ is the ion charge state and $N_i$ is the ion density. Here, $\omega_{pe}$
is the electron plasma frequency, and $\nu_{ei}^{\rm brag} $ is the
electron-ion scattering rate as defined by Braginskii.\cite{Braginskii1965}
However, over the time scales that EPWs drive decay instabilities,
\textit{i.e.}, time scales of the order of $ \gtrsim 10^{3} \omega_{pe}^{-1} $
or larger, these loss rates may be significant \cite{Brunner15,Berger2015}.
In addition, particles trapped by large amplitude waves can be scattered out
of resonance by both pitch-angle scattering and thermalization in velocity
space. Such processes have been estimated to limit the lifetime of BGK-type
equilibria.\cite{Berger2013} In multiple spatial dimensions and for waves with
a finite transverse width envelope, transverse convective losses must also be
considered.\cite{Strozzi2010, Winjum2013} The most important collisional effects are often
associated with pitch-angle scattering, whose primary effect is to change the
direction of the particle's velocity with negligible energy loss. Such
redirected particles may carry energy away from a spatially localized wave. In
this manuscript, we discuss the effects of pitch-angle collisions based on
results from Vlasov-type simulations. In addition some details are provided
concerning the implementation of collisions in the 4D = 2D+2V (two
configuration space + two velocity space dimensions) Vlasov code
LOKI,\cite{banks10,banks11} which uses fourth-order-accurate, conservative,
finite-difference algorithms.

The effect of collisions on the Landau damping of EPWs has been the subject of
a number of publications since the 1960s. A recent analytic treatment
concerned the effect on the Landau resonance of weak electron-ion collisions
in 3D velocity space \cite{Callen2014} but earlier analytic studies were also
done with 1D velocity diffusion operators.\cite{Su1968, Auerbach77, Zheng2013}
These studies did not consider the direct collisional damping of EPWs,
analogous to inverse bremsstrahlung, and easily obtained from a fluid
description by including drag in the electron momentum evolution
equations. However, Brantov\cite{Brantov2004, Brantov12} considered the more
general problem of EPW damping by using a Legendre expansion of the linearized
Fokker-Planck equations including Landau collision operators. This work,
similar to previous treatments of ion acoustic waves,\cite{Epperlein1992,
  Berger2005} found the combination of Landau and direct collisional damping
appeared as separate summable effects.

In this work, we use the 2D+2V Vlasov code LOKI, including an
electron-ion, pitch-angle collision operator, to compute the damping of
an EPW initialized with a small-amplitude density perturbation. For 
weak electron-ion collisions, the
perturbation decays exponentially in accord with the Landau damping
rate for an EPW. For larger
collision rates, we find an increase in the damping rate above the 
Landau rate because of collisional damping. Moreover, we find the 
damping directly attributable to collisions is about a factor of two
smaller than we obtain from a linearized set of fluid equations with 
electron-ion momentum exchange. To our knowledge, this result has not
been published before. 

The remainder of this manuscript is organized as follows. The basic
governing equations and discretization scheme used in the LOKI
code are briefly reviewed in Section~\ref{sec:LOKI}.  
That section presents the collision operator
considered in this work: A Lorentz pitch-angle scattering operator
restricted to two-dimensional velocity space. Using the physically
correct 3D collision operator is computationally prohibitive as it
would require extending LOKI from a 2D to a 3D velocity grid. 
Detailed exposition of the numerical methods 
are given in Appendix~\ref{sec:discretization}.
The effects of collisions on EPW damping 
are discussed for a range of $ k \lambda_{De} $ and $ \nuth / \omega_{pe}$ (see Table
\ref{linear mode characteristics}) in Section~\ref{sec:damping} in the
context of standing waves initiated with a density perturbation. 
Here, $k$ stands for the wave number and $\lambda_{De}$ for the electron Debye length.  
In Sec.~\ref{damping:theory}, a set of linearized kinetic equations in 3D spherical velocity
coordinates ( Sec.~\ref{Legendre Polynomial decomposition}) and
in 2D polar velocity coordinates (Sec.~\ref{2D Fourier decomposition})
are presented for benchmarking LOKI simulation results. With use of this set of linearized
equations, the differences between 2D and 
3D collisional velocity diffusion on the EPW damping are examined in Sec~\ref{damping:sims}
and found to be modest. Also in this Section, the LOKI simulation results for the 
EPW frequency and damping rates are shown to be in excellent agreement with the 2D linearized set.
The collisional damping of EPWs is obtained from a set of fluid equations for the density, flow
velocity, and temperature in Sec.\ref{collision damping:theory} and shown to be about twice larger 
than obtained from the kinetic equations in the strongly collisional limit. 

In addition to the
EPW mode, additional weakly damped, zero-frequency 'entropy' modes are observed in
the solutions to linearized equations and in LOKI simulations. 
These modes are not the primary interest of the current work but their presence
complicates the extraction of the EPW frequency and damping rates. 
As a result, Sec.\ref{sec:entropy} discusses these modes in relation to mitigating their effect
on computing EPW frequencies and damping rates. 

The application of LOKI to a nonlinear problem is addressed in 
Sec.~\ref{sec:NonlinearEPW} where the effect of pitch angle collisions
on the trapping of electrons in a large-amplitude EPW is studied.
Some concluding remarks are made in Section~\ref{sec:conclusions}.  Finally
the implementation of the pitch angle collision operator in the LOKI code is
discussed in detail in Appendix~\ref{sec:discretization}. It should be
noted that although the results presented in this paper concern only
electron-ion collisions, the collision algorithm in LOKI is implemented with
the ability to study collisional ion dynamics as well, for an arbitrary number
of ion species.  In Appendix~\ref{sec:discretization}, we also examine the effect
that a cap on the pitch-angle collision rate (used in the numerical
implementation of the collision operator) has on the results. The effect is insignificant
if the cap is applied only at small velocity.
In Appendix~\ref{sec:diffusivedamping}, we find an approximate solution to the
linear dispersion relation including pitch-angle collisions valid for $k
\lambda_{ei} \ll 1$. We solve that dispersion relation in 2D and 3D and find a collisional
EPW damping rate that agrees well with the Vlasov simulations. 

\section{Vlasov Equation with Collisions in 2D+2V}
\label{sec:LOKI}
We briefly describe the Vlasov-Poisson system of equations that LOKI
solves for an unmagnetized plasma in two space and two
velocity dimensions.  The evolution with respect to time $t$ of the
electron distribution function $f(x, y, v_x, v_y, t)$ is described by
the Fokker-Planck equation, {\it i.e.} the Vlasov equation including
collisional effects:
\begin{equation}
\label{fokkerplanck}
{\pt{f}+v_x\px{f}+v_y\py{f}}-\frac{e}{m_e}\left(E_x\pvx{f}+E_y\pvy{f}\right)=-C_{ei} f,
\end{equation}
with $-e$ the elementary electric charge of the electron and $m_e$ its
mass. The form of the electron-ion collision operator $C_{ei} f$ on
the right-hand side of Eq. (\ref{fokkerplanck}) will be discussed in
detail in the next sub-section. The LOKI code~\cite{banks10,banks11}
discretizes the Vlasov equations in four dimensional phase space
corresponding to the two-dimensional
configuration space $(x, y)$ and the two-dimensional velocity space
$(v_x, v_y)$.  The electric field components are given through
derivatives of the electric potential, $\phi(x,y,t)$:
\[
E_x = -\px{\phi}, \qquad E_y = -\py{\phi},
\]
which is itself determined by Poisson's equation,
\begin{equation}
\label{poisson Eq.}
\frac{\partial^2 \phi}{\partial x^2}+\frac{\partial^2 \phi}{\partial y^2} = 
   4\pi e \left( \int_{-\infty}^{\infty}\int_{-\infty}^{\infty}f_e\,dv_x dv_y - \sum_iN_iZ_i \right),
\end{equation}
where the sum is over all ion species of density $N_i$ and charge $Z_i e$ with
overall charge neutrality assumed, $N_e = \sum_i N_i Z_i$, $N_e$ standing for
the average electron density. Note that the Poisson equation (\ref{poisson
  Eq.}) has been written in cgs units (this convention is kept throughout this
paper) and that ions are considered as a fixed homogeneous, neutralizing
background, which is a very good approximation for the simulations of EPWs that
will be discussed.

%
We consider here the specific case of electron-ion pitch angle scattering, 
described by the Lorentz operator given by
\begin{eqnarray}
  \label{Lorentz coll. op.}
  C_{\rm ei}\, f 
  &=& 
  -\nu_{\rm ei, th}\,\vthe^3\,\frac{\partial}{\partial\v}\cdot
  \mathbf{U}\cdot\frac{\partial f}{\partial\v}, \\
  \nu_{\rm ei, th} &=& 2\pi \frac{Z e^4 N_e \log{\Lambda}}{m_e^2 \vthe^3},
\end{eqnarray}
where $\nu_{\rm ei, th}$ stands for the thermal electron-ion collision frequency
\footnote{$\nu_{ei}$ is defined in Reference \cite{Brantov12} as twice
  the value in Eq. \ref{Lorentz coll. op.}}, and the tensor
$\mathbf{U}$ is defined by
\begin{equation}
  \label{tensor U}
  \mathbf{U}(\v) 
  = \frac{1}{v^3} 
  \left(
  v^2\mathbf{I} - \v:\v
  \right)
  = \frac{1}{v}\, 
  \mathbf{P}_\perp,
\end{equation}
with $\mathbf{I}$ the identity tensor and $\mathbf{P}_\perp =
\mathbf{I} - \v:\v/v^2$ the projection tensor on the plane
perpendicular to the velocity $\v$.

In three-dimensional velocity space, spherical velocity variables $(v, \theta,
\phi)$ are the natural coordinates for representing the rotationally invariant
collision operator $C_{ei}$, with $v = |\v|$, $\theta$ the polar angle, and
$\phi$ the azimuthal angle. The collision operator $C_{ei}$ can indeed be
written:
\begin{equation}
  \label{Lorentz coll. op. in spherical coords.}
  C_{ei} = \nu_{\rm ei}(v)\,L^2,
\end{equation}
where $L^2$ is only dependent on the spherical-coordinate angles $(\theta, \phi)$:
\begin{flalign}
  \nonumber
  L^2 
  & = -\left[
    \frac{1}{\sin\theta}
    \frac{\partial}{\partial\theta}
    \left(
    \sin\theta \frac{\partial}{\partial\theta}
    \right)
    + 
    \frac{1}{\sin^2\theta}
    \frac{\partial^2}{\partial\phi^2}
    \right] \\
  \label{L^2, spherical variables}
  & = -\left\{
  \frac{\partial}{\partial\xi}
  \left[
    (1-\xi^2)
    \frac{\partial}{\partial\xi}
    \right]
  + 
  \frac{1}{1-\xi^2}
  \frac{\partial^2}{\partial\phi^2}
  \right\},
\end{flalign}
with the pitch angle variable $\xi = \cos(\theta)$ and the
velocity-dependent electron-ion collision frequency
\begin{equation}
  \label{e-i collision freq.}
  \nu_{\rm ei}(v) = \nu_{\rm ei, th}
  \left(
  \frac{\vthe}{v}
  \right)^3.
\end{equation}
The eigenfunctions of the collision operator (\ref{Lorentz coll. op. in
  spherical coords.}) are of the form:
\begin{equation}
  \label{eigenmode of C_ei}
  f_\lambda(v, \theta) = \delta(v-v_0)\,Y^m_l(\theta, \phi),
\end{equation}
associated to the eigenvalues
\begin{equation}
  \label{eigenvalue of C_ei}
  \lambda = \nu_{\rm ei}(v_0)\,l(l+1).
\end{equation}
In Eq. (\ref{eigenmode of C_ei}), $\delta(v-v_0)$ stands for the Dirac delta
function centered at an arbitrary velocity amplitude $v_0\in\mathbb{R}_+$ and
$Y^m_l(\theta, \phi)$ represents a spherical harmonic of degree
$l\in\mathbb{N}$ and order $m\in\mathbb{Z}$, $m=-l, -l+1, \ldots, +l$. The spherical harmonics are given by
\[
Y^m_l(\theta, \phi) = P^m_l(\cos\theta)\, e^{i m\phi},
\]
where the functions $P^m_l(\xi)$ are the associated Legendre polynomials of
the first kind. In the LOKI model considered in this work, velocities are restricted to the
two-dimensional space $(v_x, v_y)$, corresponding to the two Cartesian
configuration space dimensions $(x, y)$. Reduction from the full
three-dimensional velocity space to two-dimensional velocity space is
necessary because discretization in six-dimensional [or even five-dimensional
  (2D+3V)] phase space remains computationally prohibitive. The natural
coordinates for representing the collision operator (\ref{Lorentz coll. op.})
restricted to this two-dimensional velocity space are polar coordinates $(v,
\theta)$, with $v=|\v|$ the velocity amplitude and the poloidal angle $\theta$
such that $\cos\theta = v_x/v$ and $\sin\theta = v_y/v$. In these coordinates
the collision operator (3) takes the particularly simple form:
\begin{equation}
  \label{Lorentz coll. op. in polar coords.}
  C_{ei} = -\nu_{\rm ei}(v)\frac{\partial^2}{\partial\theta^2},
\end{equation}
with velocity dependent electron-ion collision frequency given by Eq. (\ref{e-i collision freq.}).
The eigenfunctions of the collision operator (\ref{Lorentz
  coll. op. in polar coords.}) are of the form
\begin{equation}
  \label{eigenmode of C_ei, restricted}
  f_\lambda(v, \theta) = \delta(v-v_0)\,e^{im\theta},
\end{equation}
associated to the eigenvalues
\begin{equation}
  \label{eigenvalue of C_ei, restricted}
  \lambda = \nu_{\rm ei}(v_0)\,m^2,
\end{equation}
where $m \in \mathbb{Z}$ is an arbitrary integer and $\delta(v-v_0)$ again
stands for the Dirac delta function centered at an arbitrary velocity
amplitude $v_0\in\mathbb{R}_+$.

In LOKI, velocity is however represented using the Cartesian coordinates
$(v_x, v_y)$, a more natural representation for the collisionless advection
dynamics defined by the left hand side of Eq. (\ref{fokkerplanck}). The
operator (\ref{Lorentz coll. op. in polar coords.})  must therefore also be
expressed in the same coordinates $(v_x, v_y)$, which is a less natural choice
for this collision term.  Furthermore, in order to retain the strictly
conservative formulation of LOKI, we express the two-dimensional pitch angle
scattering operator in the conservative form:
\begin{align}
  \nonumber
  C_{ei}f
  = & \frac{\partial}{\partial\v}\cdot\vec{\Gamma}
  = \frac{\partial}{\partial v_x} \Gamma_x
  + \frac{\partial}{\partial v_y} \Gamma_y \\
  \nonumber
  = & \,
  \left\{
  \frac{\partial}{\partial v_x}
  \left[
    \nu_{\rm ei}(v)
    v_y
    \left(
    v_x\frac{\partial f}{\partial v_y}
    -v_y\frac{\partial f}{\partial v_x}
    \right)
    \right] 
  \right. \\
  \label{Lorentz coll. op. in Cartesian coords.}
  & \hspace{8.mm}
  \left.
  - \frac{\partial}{\partial v_y}
  \left[
    \nu_{\rm ei}(v)
    v_x
    \left(
    v_x\frac{\partial f}{\partial v_y}
    -v_y\frac{\partial f}{\partial v_x}
    \right)
    \right]
  \right\}.
\end{align}
Note that equation (\ref{Lorentz coll. op. in Cartesian coords.}) can also be
written in the following non-conservative form:
\[
 C_{ei}f
=
-\nu_{\rm ei}(v)
\left(
v_x\frac{\partial}{\partial v_y}
    -v_y\frac{\partial}{\partial v_x}
\right)^2 f,
\]
obviously equivalent to (\ref{Lorentz coll. op. in polar coords.}) given that
$\partial/\partial\theta = v_x\partial/\partial v_y -v_y\partial/\partial
v_x$.  

Discretization of the collision operator (\ref{Lorentz coll. op. in Cartesian
  coords.}) using a higher order finite difference scheme is discussed in
Appendix~\ref{sec:discretization}. This appendix also explains how the
velocity dependent collision frequency $\nu_{ei}(v)$ is slightly modified in
LOKI for practical reasons. In particular, $\nu_{ei}(v)$ is capped at very low
velocities: $\nu_{ei}(v) \equiv \nu_{ei}(\bar{v}) = \nu_{ei, \max}$ for
$v<\bar{v}\ll\vthe$ [see Eq. (A7)]. Without this cap, the collision frequency
diverges for velocities approaching zero, $\nu_{ei}(v)\sim v^{-3} \to \infty$
as $v \to 0$, and the time step required for the explicit time integration
scheme implemented in LOKI would be impractically small. In
Appendix~\ref{sec:discretization}, results relevant to those discussed in the
main part of this paper illustrate that a cap set as low as $\nu_{ei, \max} =
10\,\nu_{ei, th}$, corresponding to $\bar{v}/\vthe = 10^{-1/3} \simeq 0.464$,
has no discernable effect on the results for EPW damping. 
Furthermore, as $\nu_{ei}(v)\sim v^{-3}$, the collision rate 
naturally goes to zero as $v \to \infty$. However,
for any {\it finite} maximum velocity of the grid, collisionality is non-zero
and the collision operator therefore changes the nature of the equation from
hyperbolic ({\it i.e.} purely advective) to parabolic, with undesirable
consequences when implementing 
the boundary conditions. For that reason, the
velocity dependent collision frequency $\nu_{ei}(v)$ is furthermore modified
to smoothly go to zero for velocities $v_c < v < v_{\max}$, with $v_c$ short
of the maximum value $v_{\max}$ of the grids along $v_x$ and $v_y$ [see as
  well Eq. (A7)].  Ensuring that $\nu_{ei}(v_{max}) = 0$ thus restores the
hyperbolic boundary conditions, again with no effect on the results if
$v_{max}$ is large enough. The results presented in Sec.~\ref{sec:damping}
were carried out with $\nu_{ei, \max} = 100\,\nu_{ei, th}$,
which corresponds to $\bar{v}/\vthe = 100^{-1/3} \simeq 0.215$ and with $v_{max}/\vthe
= 7$ for $k \lambda_{De} = 0.3$ and $0.425$ but with $v_{max}/\vthe = 10$ for $k
\lambda_{De} = 0.2$.

\section{Collisional effects on Electron Plasma wave damping}
\label{sec:damping}

When considering the Lorentz electron-ion pitch angle scattering collision
operator restricted to two-dimensional velocity space instead of the full
three-dimensional space, the question naturally arises how this approximation
affects the collisional processes. To address this issue, we consider as a
test problem the effect of varying collisionality on the damping, both Landau
and collisional, of linear EPWs.
%
%
\subsection{Linearized Kinetic Equations in Spherical and Polar Velocity Variables}
\label{damping:theory}

In the following, we decompose the linearized
Vlasov-Poisson system including the pitch angle scattering operator in
three-dimensional velocity space, as described by Eqs. (\ref{Lorentz
  coll. op. in spherical coords.}) and (\ref{L^2, spherical variables}), using
a Legendre polynomial representation (see {\it e.g.}
Ref. \cite{Epperlein1992}). The Legendre polynomials (more generally
spherical harmonics) are eigenfunctions of the operator $L^2$ appearing in the
collision operator $C_{ei}$ given by Eq. (\ref{Lorentz coll. op. in spherical
  coords.}).  In two-dimensional velocity space, a similar system of equations 
Eq. (\ref{Lorentz coll. op. in polar coords.}) is appropriate. To this end we make use of a
Fourier mode decomposition with respect to the polar angle $\theta$, as such
modes correspond to eigenfunctions of the restricted operator [see
  Eq. (\ref{eigenmode of C_ei, restricted})]. The relations obtained for
the linearized Vlasov-Poisson system are the natural coordinates with respect to
the collisional dynamics and provide an alternative set of equations that are
also solved numerically and compared to the LOKI simulation results 
for benchmarking purposes 
in Sec. \ref{damping:sims}. Furthermore, these relations 
enable the consideration of approximate or limiting cases that provide insight into the interpretation of
the collisional effects on EPWs, as will be done in Secs. \ref{interpretation}
and \ref{sec:entropy} as well as Appendix \ref{sec:diffusivedamping}.

The full electron distribution $f$ is decomposed into a Maxwellian background $f_M(v)$ and a
small fluctuating part $\delta f$:
\[
f(\x, \v, t) = f_M(v) + \delta f(\x, \v, t).
\]
such that the linearized Vlasov-Poisson system of equations including electron-ion
collisions becomes:
\begin{gather}
  \label{Linearized Vlasov Eq. with coll.}
  \frac{\partial \delta f}{\partial t} 
  + v_x \frac{\partial \delta f}{\partial x} 
  + \frac{\partial \phi}{\partial x}\frac{\partial f_M}{\partial v_x} 
  = -C_{\rm ei} \delta f, \\
  \label{Poisson Eq. with df source}
  \frac{\partial^2\phi}{\partial x^2}
  =  \int \delta f\,d^Dv,
\end{gather}
with use of the normalizations $t\,\omega_{\rm pe} \to t$, $x/\lambda_{\rm De}
\to x$, $\v/\vthe \to \v$, $e\,\phi\,/T_e \to \phi$, and $f\vthe^D/N\to f$,
with $D=2,3$ as appropriate to the number of velocity dimensions.
%
%
\subsubsection{Scattering in 3D and Legendre Polynomial decomposition}
\label{Legendre Polynomial decomposition}
For scattering in three-dimensional velocity space, $C_{\rm ei} = \nu_{\rm
  ei}(v)\,L^2$, as given by Eqs. (\ref{Lorentz coll. op. in spherical
  coords.})-(\ref{L^2, spherical variables}).
As the unperturbed system is translationally invariant in the $x$-direction,
the linear analysis can be reduced to the study of independent Fourier modes
with wavenumber $k$. Furthermore, spherical harmonics $Y_l^m(\theta, \phi)$
are eigenmodes of the Lorentz collision operator in the spherical angle space
$(\theta, \phi)$.  Thus a spherical harmonic representation for the velocity
angle dependence of the fluctuating part $\delta f$ of the distribution is
chosen.  These angular modes, however, become coupled through the advective
part of the Vlasov equation, i.e. the left hand side of Eq. (\ref{Linearized
  Vlasov Eq. with coll.}). As the system is azimuthally symmetric (having
aligned the zenith direction of the spherical velocity variables with the
Cartesian direction $v_x$), the spherical harmonic representation reduces to a
Legendre polynomial decomposition with respect to the pitch angle $\xi =
\cos\theta$ dependence:
\begin{equation}
  \label{Legendre polynomial decomposition of df}
  \delta f(x, v, \xi, t) = e^{ikx}\,\sum_{l=0}^{\infty} \delta f_l (v, t) P_l(\xi),
\end{equation}
Inserting
(\ref{Legendre polynomial decomposition of df}) into Eq. (\ref{Linearized
  Vlasov Eq. with coll.}), projecting this equation onto the various Legendre
polynomials using orthogonality and recurrence relations ($\delta_{l\,l'}$ stands for the
Kronecker delta) we obtain 
\begin{gather}
  \label{Fokker-Planck, Legendre component l=0}
  l = 0: \hspace{1.cm}
  \frac{\partial\, \delta f_0}{\partial t} 
  + \frac{ikv}{3}\delta f_1
  =0,\\
  \label{Fokker-Planck, Legendre component l=1}
  l = 1: \hspace{1.cm}
  \frac{\partial\, \delta f_1}{\partial t} 
  + ikv\,\left(\delta f_0
  + \frac{2}{5}\delta f_2
  \right)
  - ikv\,\phi\,f_M
  =
  - 2\,\nu_{\rm ei}(v)\,\delta f_1, \\
  \label{Fokker-Planck, Legendre component l>=2}
  l \ge 2: \hspace{1.cm}
  \frac{\partial\, \delta f_l}{\partial t} 
  + ikv\,\left(
  \frac{l}{2l-1}\delta f_{l-1}
  + \frac{l+1}{2l+3}\delta f_{l+1}
  \right)
  =
  - \nu_{\rm ei}(v)\,l(l+1)\,\delta f_l,
\end{gather}
while the Poisson equation (\ref{Poisson Eq. with df source}) becomes:
\begin{equation}
  \label{Poisson Eq. with df0 source}
  k^2 \phi = - 4\pi\int_0^{+\infty} v^2dv\, \delta f_0.
\end{equation}
The time evolution of the linear system of Eqs. (\ref{Fokker-Planck, Legendre
  component l=0})-(\ref{Poisson Eq. with df0 source}) is solved numerically.
For this, the velocity amplitude $v$ was discretized over an interval $[0,
  v_{\rm max}]$ with an equidistant mesh $\{v_j\}_{j=1, \ldots, n_v}$ and the
integral in (\ref{Poisson Eq. with df0 source}) estimated with a composite
trapezoidal rule. After inserting (\ref{Poisson Eq. with df0 source}) into
(\ref{Fokker-Planck, Legendre component l=1}), the system (\ref{Fokker-Planck,
  Legendre component l=0})-(\ref{Fokker-Planck, Legendre component l>=2})
defines a system of linear first order ordinary differential equations for the
coefficients $\delta f_{l, j}(t) = \delta f_l(v_j, t)$, $l = 0,\ldots,
L_{\max}$, $j = 1, \ldots, n_v$, which is solved with the initial condition
\begin{equation*}
  \label{initial condition on df0}
  \delta f_l(v, t=0) = \delta n f_M(v)\, \delta_{l, 0},
\end{equation*}
corresponding to a sinusoidal density perturbation of the
Maxwellian background velocity distribution with relative amplitude $\delta
n$. Note that for scattering in three-dimensional velocity space, the
normalized Maxwellian distribution is chosen as
\begin{equation}
  \label{normalized Maxwellian in D dimensions}
  f_M(v) 
  = 
  \frac{1}{(2\pi)^{D/2}}
  \exp\left(-\frac{v^2}{2}\right),
\end{equation}
with $D=3$.
%
%
\subsubsection{Scattering in 2D and Fourier decomposition}
\label{2D Fourier decomposition}
Here, the same linearized Vlasov-Poisson system with collisional dynamics 
as given by Eqs. (\ref{Linearized Vlasov Eq. with coll.})-(\ref{Poisson Eq. with
  df source}) is analyzed in two-dimensional velocity space given by the operator $C_{ei} = -\nu_{\rm
  ei}(v)\partial^2/\partial\theta^2$ as already defined in Eq. (\ref{Lorentz coll. op. in polar coords.}). In this case
a Fourier mode decomposition with respect to the
polar angle $\theta$ is used, as Fourier modes are the eigenmodes of the
2D scattering operator:
\begin{equation}
  \label{Polar Fourier mode decomposition of df}
  \delta f(x, v, \theta, t) 
  = 
  e^{ikx}\,\sum_{m=-\infty}^{+\infty} 
  \delta f_m (v, t) \exp(im\theta).
\end{equation}
The complex Fourier representation (\ref{Polar Fourier mode decomposition of df}) is
 equivalent to the sine-cosine decomposition:
\begin{equation}
  \label{sine-cosine decomposition of df}
  \delta f(x, v, \theta, t) 
  = 
  e^{ikx}\,
  \left[
    \sum_{m=0}^{+\infty} 
    \delta\!f_{c,m} (v, t) \cos(m\theta)
    +
    \sum_{m=1}^{+\infty} 
    \delta\!f_{s,m} (v, t) \sin(m\theta)
    \right].
\end{equation}
The following relations between these two representations are obtained:
\begin{gather*}
\delta f_0  = \delta\!f_{c,0} \\
\delta f_m = \frac{\delta\!f_{c,m}}{2} + \frac{\delta\!f_{s,m}}{2i}, 
\hspace{0.5cm}
\text{and}
\hspace{0.5cm}
\delta f_{-m} = \frac{\delta\!f_{c,m}}{2} - \frac{\delta\!f_{s,m}}{2i},
\hspace{0.5cm}
\text{for}
\hspace{0.5cm}
m \ge 1.
\end{gather*}
It will be shown in the following that representation (\ref{sine-cosine
  decomposition of df}) is convenient for highlighting symmetry properties of
the Vlasov-Poisson system (\ref{Linearized Vlasov Eq. with
  coll.})-(\ref{Poisson Eq. with df source}). Inserting (\ref{sine-cosine
  decomposition of df}) into (\ref{Linearized Vlasov Eq. with coll.}) and
projecting this equation onto the different sine-cosine modes using the
orthogonality relations yields:
\begin{flalign}
  \label{Fokker-Planck, cosine component m=0}
  m = 0: \hspace{4.5cm}
  \frac{\partial\, \delta\!f_{c,0}}{\partial t} 
  + ikv\frac{\delta\!f_{c,1}}{2}
  & = 0,\\
  \label{Fokker-Planck, cosine component m=1}
  m = 1: \hspace{0.5cm}
  \frac{\partial\, \delta\!f_{c,1}}{\partial t} 
  + ikv\,
  \left(
  \delta\!f_{c,0} + \frac{\delta\!f_{c,2}}{2}
  \right)
  - ikv\,\phi\,f_M
  & =
  - \nu_{\rm ei}(v)\,\delta\!f_{c,1}, \\
  \label{Fokker-Planck, sine component m=1}
  \frac{\partial\, \delta\!f_{s,1}}{\partial t} 
  + ikv\,
  \frac{\delta\!f_{s,2}}{2}
  & =
  - \nu_{\rm ei}(v)\,\delta\!f_{s,1}, \\
  \label{Fokker-Planck, cosine component m>=2}
  m \ge 2: \hspace{1.3cm}
  \frac{\partial\, \delta\!f_{c,m}}{\partial t} 
  + ikv\,
  \left(
  \frac{\delta\!f_{c,m-1}}{2} +   \frac{\delta\!f_{c,m+1}}{2}
  \right)
  & =
  - \nu_{\rm ei}(v)\,m^2\,\delta\!f_{c,m}, \\
  \label{Fokker-Planck, sine component m>=2}
  \frac{\partial\, \delta\!f_{s,m}}{\partial t} 
  + ikv\,
  \left(
  \frac{\delta\!f_{s, m-1}}{2} +   \frac{\delta\!f_{s,m+1}}{2}
  \right)
  & =
  - \nu_{\rm ei}(v)\,m^2\,\delta\!f_{s,m}, 
\end{flalign}
while the Poisson equation (\ref{Poisson Eq. with df source}) becomes:
\begin{equation}
  \label{Poisson Eq. with dfc0 source}
  k^2 \phi = - 2\pi\int_0^{+\infty} \hspace{-0.4cm}v\,dv\, \delta\!f_{c,0}.
\end{equation}
After again discretizing the velocity amplitude $v$ over an interval $[0,
  v_{\rm max}]$, the system of Eqs. (\ref{Fokker-Planck, cosine component
  m=0})-(\ref{Poisson Eq. with dfc0 source}) defines a system of linear first
order ordinary differential equations for the coefficients $\delta\!f_{c, m,
  j}(t) = \delta\!f_{c,m}(v_j, t)$, and $\delta\!f_{s, m, j}(t) =
\delta\!f_{s,m}(v_j, t)$, $m=0, \ldots, M$, $j=1, \ldots, n_v$, which, in the
same way as for system (\ref{Fokker-Planck, Legendre component
  l=0})-(\ref{Poisson Eq. with df0 source}), is solved with an initial
condition specified by a sinusoidal density perturbation of the
Maxwell-Boltzmann background velocity distribution. In the sine-cosine
representation, such an initial state reads:
\begin{equation}
  \label{initial condition on dfc0 and dfs0}
  \delta\!f_{c,m}(v, t=0) = \delta n f_M(v)\, \delta_{m, 0},
  \makebox[2.cm]{and}
  \delta\!f_{s,m}(v, t=0) = 0,
\end{equation}
which corresponds to the initial condition for the LOKI simulations in Eq.\, (\ref{initial f}).
Note that for the velocity space restricted to two-dimensions,
the Maxwellian distribution $f_M(v)$ is given by Eq. (\ref{normalized Maxwellian in
  D dimensions}) with $D=2$.

Remarkable in the system (\ref{Fokker-Planck, cosine component
  m=0})-(\ref{Poisson Eq. with dfc0 source}) is the fact that the cosine
coefficients $\delta\!f_{c,m}(v, t)$ are decoupled from the sine coefficients
$\delta\!f_{s,m}(v, t)$. Note that for the initial condition (\ref{initial
  condition on dfc0 and dfs0}), $\delta\!f_{s,m}(v, t) \equiv 0$ for all $m
\in \mathbb{N} $ and all times $t$. Consequently, only
Eqs. (\ref{Fokker-Planck, cosine component m=0}), (\ref{Fokker-Planck, cosine
  component m=1}), (\ref{Fokker-Planck, cosine component m>=2}), and
(\ref{Poisson Eq. with dfc0 source}) need be solved for the evolution of
$\delta\!f_{c, m}, m \in \mathbb{N}$, together with the initial condition
(\ref{initial condition on dfc0 and dfs0}). This system has been solved
numerically, and the results of this calculation are compared in detail for a
scan over $k \lambda_{De}$ and $\nuth/\omega_{pe}$ with the corresponding LOKI
results in the next section. These 2D velocity scattering results are also
compared to the 3D scattering results obtained from the system of
Eqs. (\ref{Fokker-Planck, Legendre component l=0})-(\ref{Poisson Eq. with df0
  source}).
%
%
\subsection{Simulation Results of Linear Electron Plasma Wave Damping}
\label{damping:sims}
In this section, simulation results obtained with the LOKI code for studying
the effect of collisions on linear damping of spatially one-dimensional EPWs
propagating in the $x$-direction with wavenumbers $k\lambda_{De} = 0.200,
0.300,$ and $ ~\rm 0.425$ are presented.  For comparison, numerical results
are also shown for the system described in Sec. (\ref{2D Fourier
  decomposition}), {\it i.e.}  essentially the same system of equations as
LOKI, in particular the electron-ion pitch angle collision operator restricted
to 2D, except that the Vlasov equation has been linearized for small
electrostatic perturbations with respect to an equilibrium state characterized
by a Maxwellian electron distribution. In the low perturbation amplitude
regime, one therefore expects the results from these two approaches to agree.
\begin{table}
\centering
\caption{Linear kinetic frequency, collisionless Landau damping rate, and
  phase velocity for EPWs with wavenumbers $k\lambda_{De} = 0.200$, $0.300$
  and $0.425$.}
\label{linear mode characteristics}
\begin{tabular}{|C{3cm}|C{3cm}|C{3cm}|C{3cm}|}
\hline\hline
$k\lambda_{De}$ & $\omega_R/\omega_{pe}$ &  $\gamma/\omega_{pe}$   &  $v_{\phi}/v_{the}$  \\
\hline
$0.200$ & 1.064 & $ 5.511 \cdot 10^{-5}$  & 5.320       \\  
$0.300$   & 1.160 & $1.262 \cdot 10^{-2}$  & $3.867$   \\  
$0.425$ & $1.318$ & $8.526 \cdot 10^{-2}$ &  $3.100$ \\  
\hline
\end{tabular}
\end{table}
In these simulations, the initial electron distribution was set to
\begin{equation}
  \label{initial f}
  f(x, v_x, v_y, t=0) 
  =
  f_M(v) + \delta f(x, v, t=0)
  =
  \left[
    1 + \frac{\delta n}{N}\cos(k x)
    \right] 
  f_M(v),
\end{equation}
corresponding to a Maxwellian distribution $f_M(v)$ with a sinusoidal density
perturbation, which evolves into a standing EPW. The distribution, $f_M(v)$ is
given by Eq. (\ref{normalized Maxwellian in D dimensions}) with $D=2$ and the
relative density perturbation in LOKI is set to $\delta n/N = 1\cdot 10^{-4}$
to ensure the simulations remain in the linear regime. Concerning the mesh
resolutions for the LOKI runs, the number of uniformally-spaced spatial grid
points over the one-wavelength long system was set to $n_x=64$, while in the
transverse direction $y$, for the spatially one-dimensional problem considered
here, the number of grid points $n_y=5$ was set (this is the minimum allowed
number in LOKI, corresponding to the stencil width of the
discretization scheme). For the velocity grids along $v_x$ and $v_y$, maximum
values $v_{x, \max} = v_{y, \max} = v_{\max} =7\,\vthe$ and uniform grid
resolutions $\Delta v_x = 2 v_{x, \max}/n_{v_x} = \Delta v_y = 2 v_{y,
  \max}/n_{v_y} = 1.09 \cdot 10^{-1}$ where $n_{v_x} = n_{v_y} = 128$ 
were considered. The corresponding results obtained from the numerical solutions to
the system (\ref{Fokker-Planck, cosine component m=0})-(\ref{Poisson Eq. with
  dfc0 source}) with polar velocity variables used maximum velocity
$v_{\max}/\vthe = 7$, velocity-amplitude grid point number $n_v = 128$, and
maximum number $M =10-20$ of polar Fourier modes, with a higher number of
polar modes required as the collision rate approached zero, $\nu_{\rm ei,
  th}/\omega_{pe}\to 0$.
\begin{figure}
  \centering
  \includegraphics[width=2.75in]{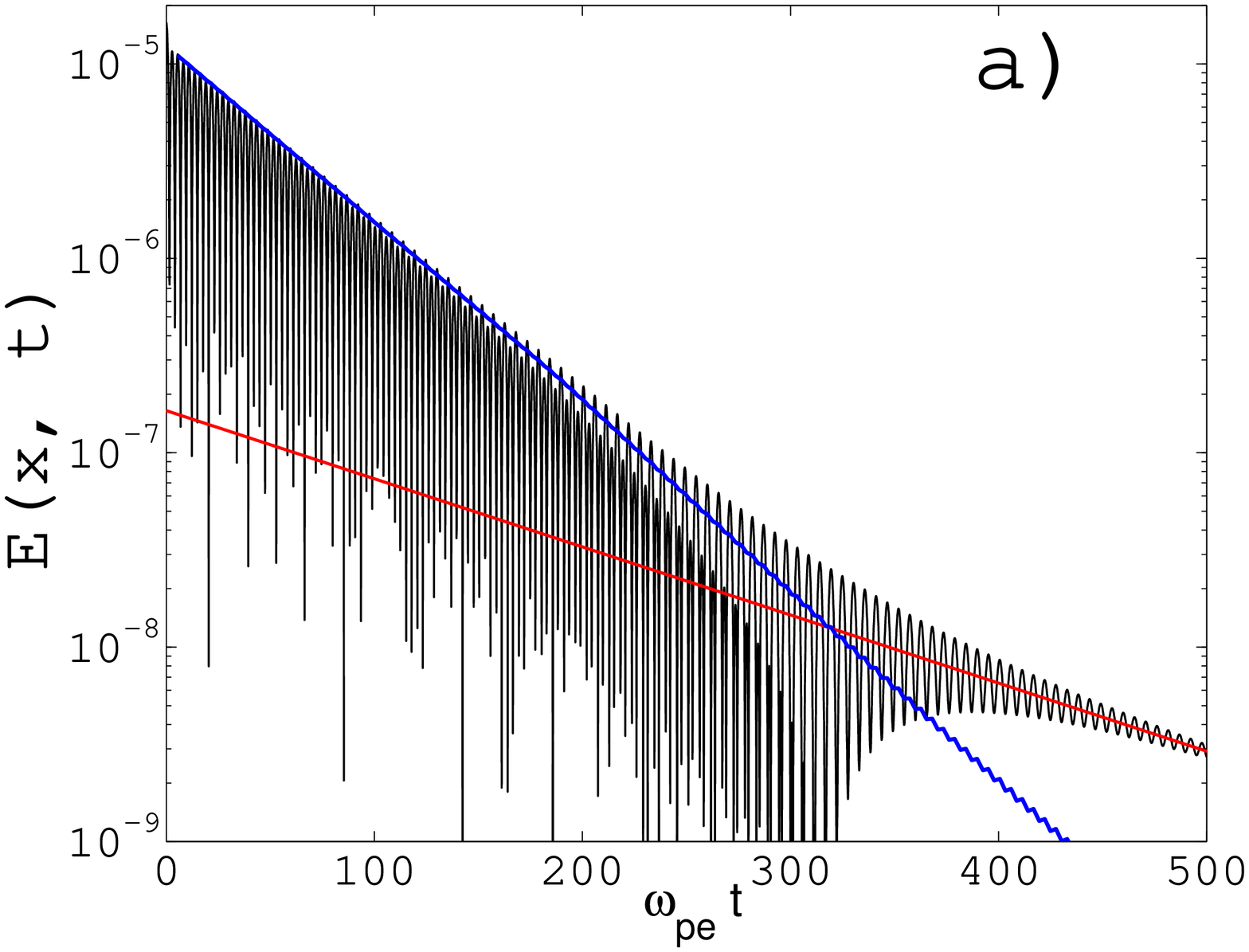}
  \includegraphics[width=2.75in]{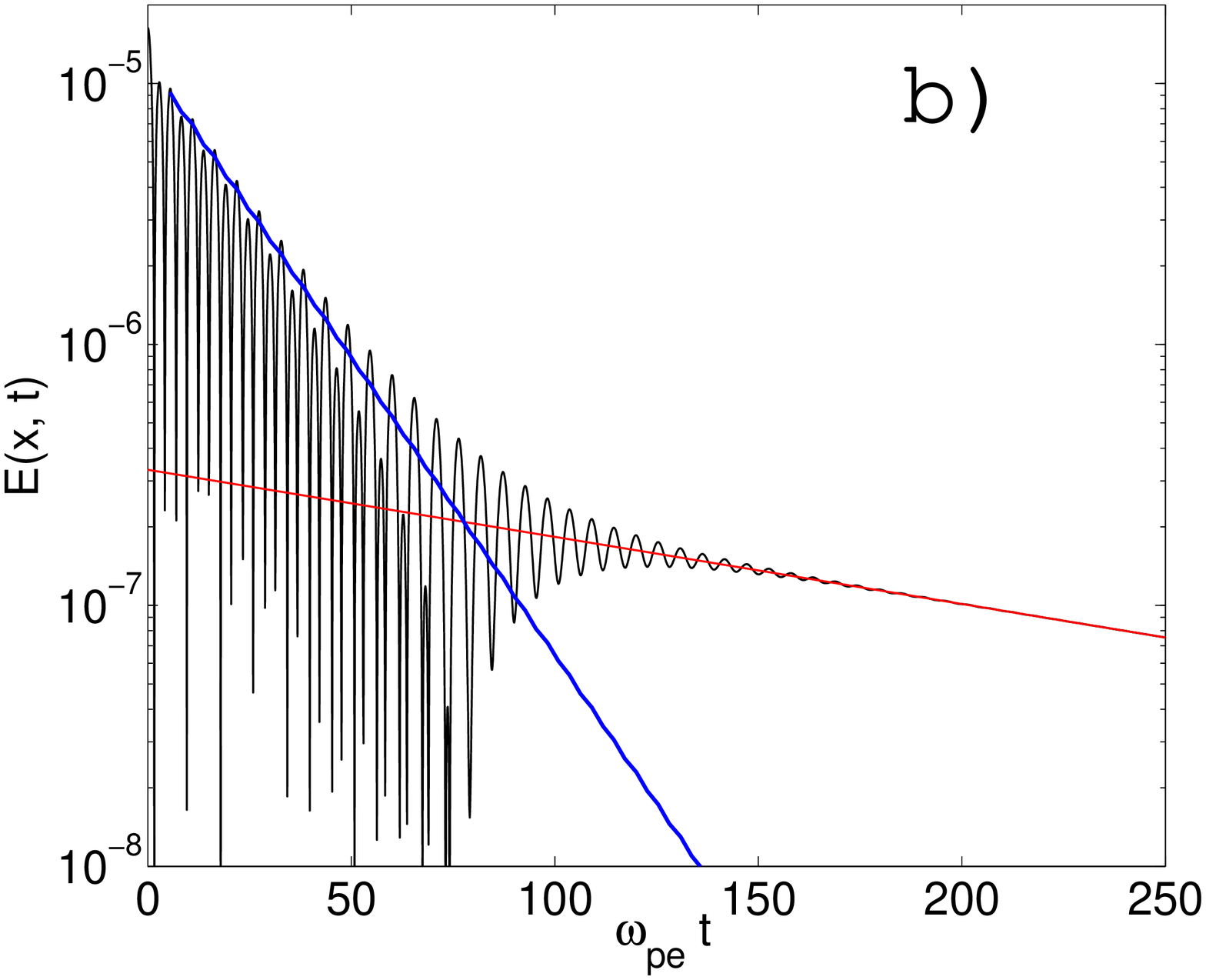}
  \includegraphics[width=2.75in]{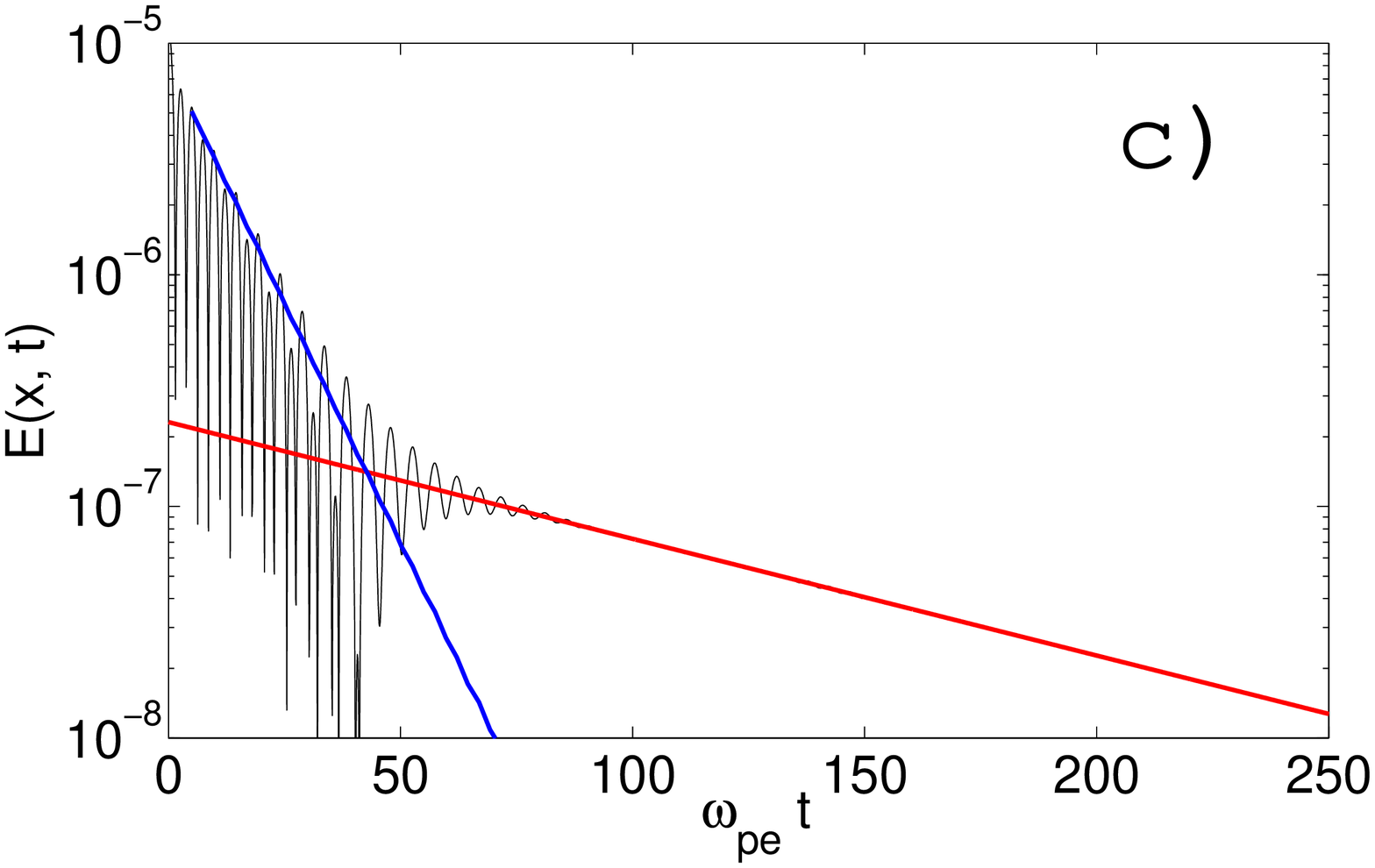}
  \includegraphics[width=2.75in]{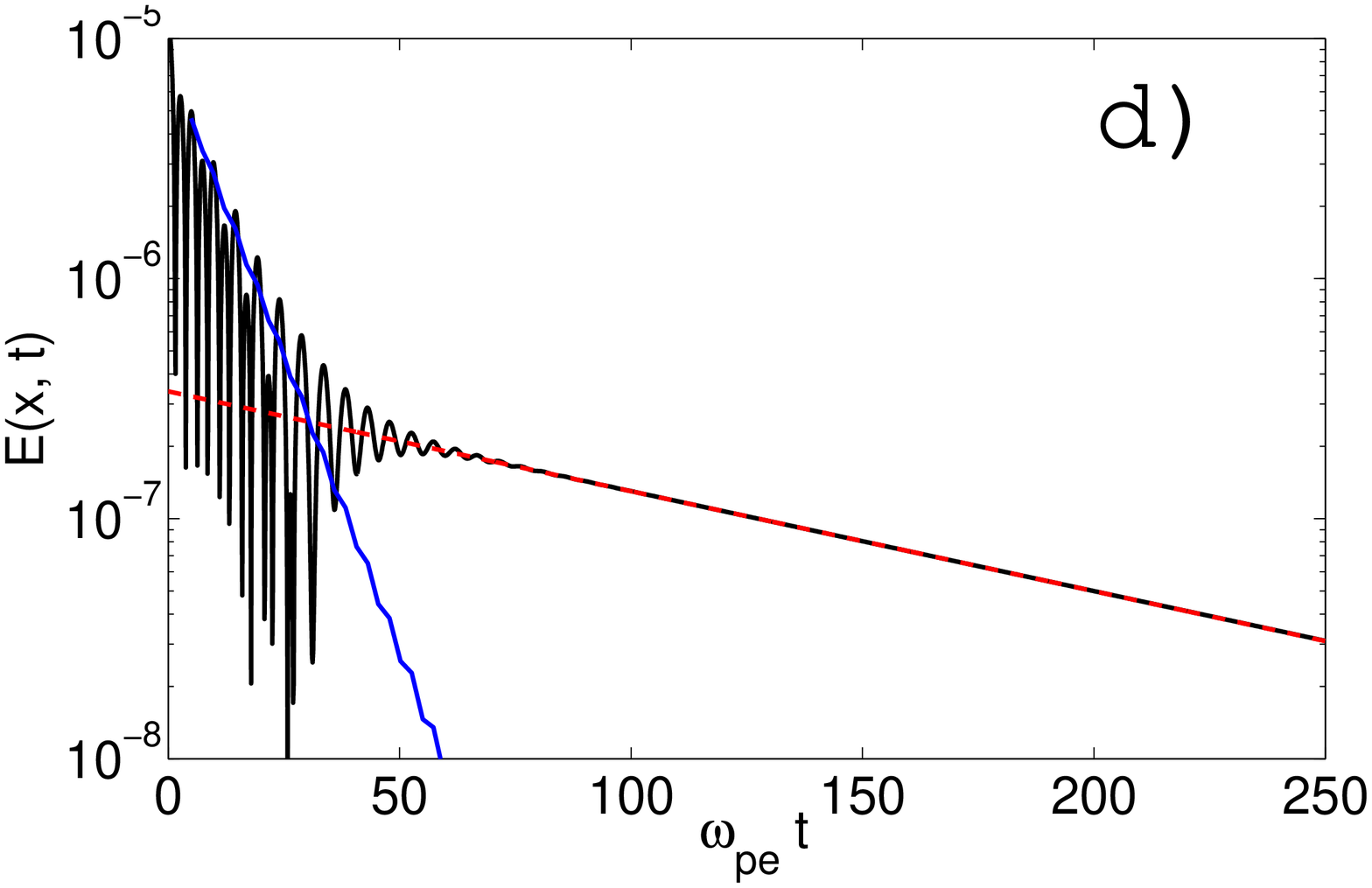}
  \caption[]{\label{Fig. lin. Landau damp., with/without coll.}  (Color
    online) LOKI simulation results for linear damping of EPWs. The amplitude
    of the perturbation's electric field $E$ is plotted on a lin-log scale as
    a function of time $\omega_{pe}\,t$ for a wave with (a) $k\lambda_{De} =
    0.300$ and $\nu_{\rm ei, th}/\omega_{pe} = 5\cdot 10^{-2}$; (b)
    $k\lambda_{De} = 0.300$ and $\nu_{\rm ei, th}/\omega_{pe} = 3\cdot
    10^{-1}$; (c) $k\lambda_{De} = 0.425$ and $\nu_{\rm ei, th}/\omega_{pe} =
    7\cdot 10^{-2}$; and (d) $k\lambda_{De} = 0.425$ and $\nu_{\rm ei,
      th}/\omega_{pe} = 2.1\cdot 10^{-1}$ 
    In all cases,
    $\nu_{\rm ei,\max}/\nu_{\rm ei, th} = 100$.  In the first phase, following
    a brief transient of $\omega_{pe} t ~\simeq 10$, one observes the damping
    of the standing EPW, whose finite real frequency leads to the oscillation
    of the amplitude. In a second phase, an initially small
      'collisional' or 'entropy' mode with weaker damping survives. Zero real
      frequency of this 'collisional' mode implies no oscillation of the
      amplitude. The red lines are an exponential decay fit to the observed
      entropy mode based on simulation data late in time. That behavior is
      extrapolated to all times and subtracted from the total field which
      allows more accurate determination of the frequency and damping of the
      EPW. The exponential decay fit to the EPW is shown by the blue
    lines.}
\end{figure}
\begin{figure}
  \centering
  \includegraphics[width=0.4\textwidth]{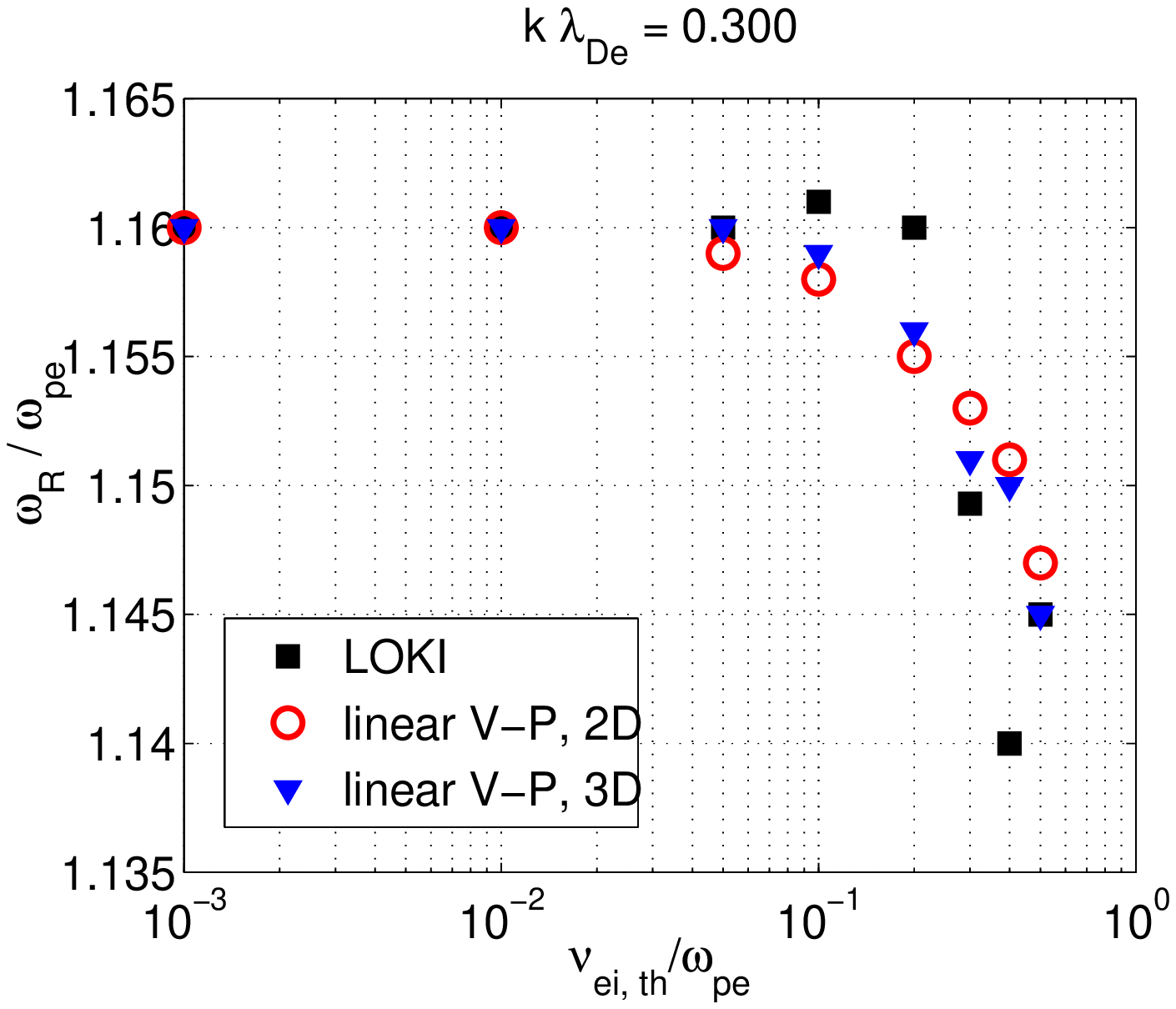}
  \includegraphics[width=0.4\textwidth]{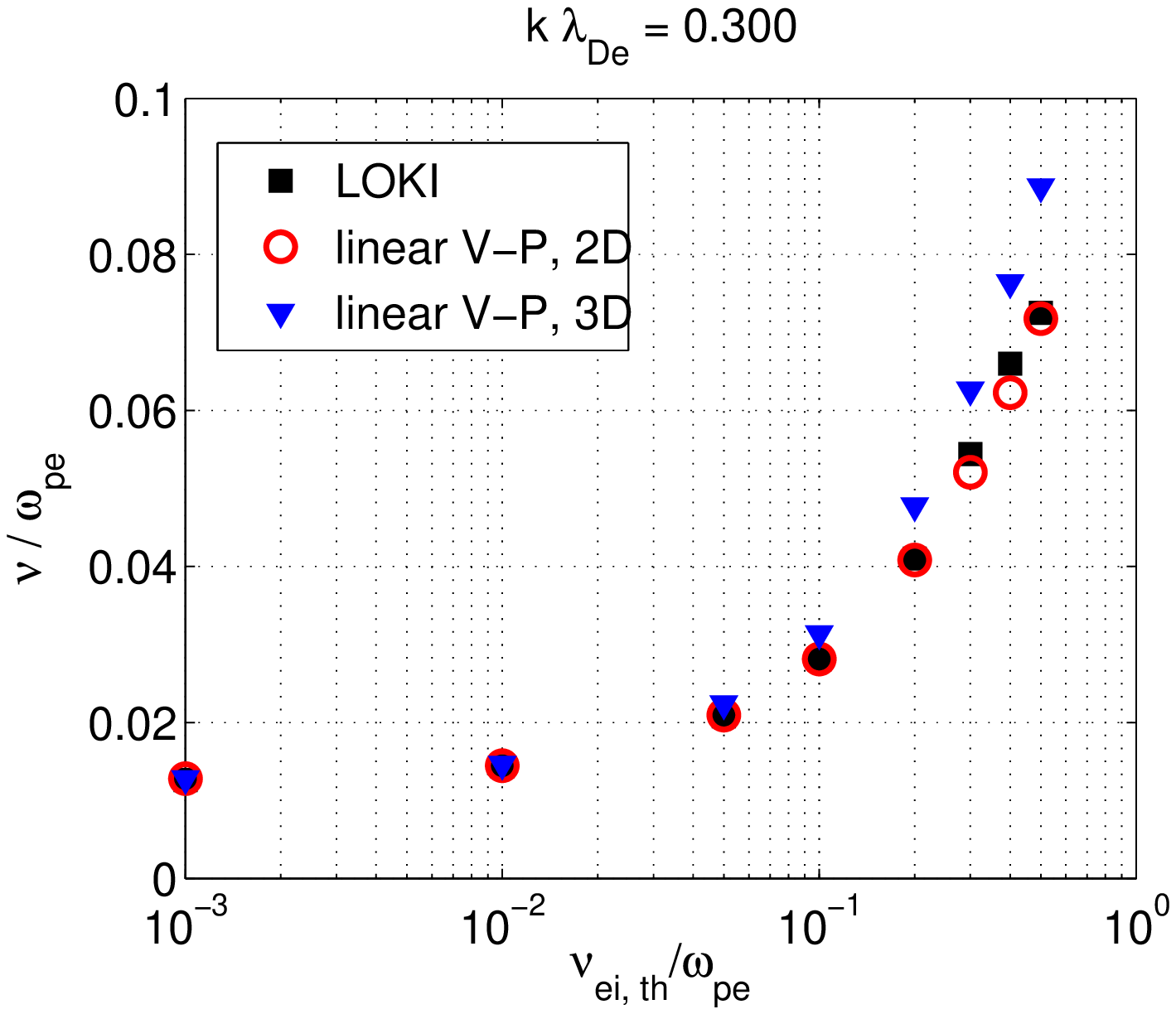}
  \caption[]{\label{Fig. Coll. effects on lin. Landau damp., 2D vs 3D,
      numax/nuth = 100} Effect of collisions on linear EPW damping for
    $k\lambda_{De} = 0.300$. LOKI simulation results (black square markers)
    are compared to the linearized Vlasov-Poisson system with electron-ion
    collisional scattering in the two- and three-dimensional velocity space
    (red circles and blue triangles respectively). For all results, the
    maximum collision rate was capped at $\nu_{\rm ei, \max}/\nu_{\rm ei, th}
    = 100$.  Plotted are (a) the real frequency, $\omega_R/\omega_{pe}$, and
    (b) the damping rate, $\nu/\omega_{pe}$, as a function of the normalized
    thermal electron-ion collision frequency $\nu_{\rm ei,
      th}/\omega_{pe}$. Note the very good agreement between the
    predictions from the linearized Vlasov-Poisson system with
    collisions in 2D velocity space and the LOKI simulations. Note also that the maximum
    relative difference on the damping rate between results for 2D and for 
    3D velocity scattering is only $20\%$ and occurs at very high collisionality.}
\end{figure}
Figures \ref{Fig. lin. Landau damp., with/without coll.}.a and
\ref{Fig. lin. Landau damp., with/without coll.}.b show time traces from LOKI
simulations of the amplitude of the EPW for $k \lambda_{De} = 0.300$, while
Figs. \ref{Fig. lin. Landau damp., with/without coll.}.c and
\ref{Fig. lin. Landau damp., with/without coll.}.d show time traces for $k
\lambda_{De} = 0.425$.  Figures \ref{Fig. lin. Landau damp., with/without
  coll.}.a to \ref{Fig. lin. Landau damp., with/without coll.}.d have $\nuth =
5\cdot 10^{-2}$, $3\cdot 10^{-1}$, $7\cdot 10^{-2}$ and $2.1\cdot 10^{-1}$
respectively. In all cases, the velocity dependent collisionality
$\nu_{ei}(v)$ has been capped at $\nu_{\rm ei, \max} = \nu_{ei}(\bar{v}) =
100\,\nu_{\rm ei, th}$, corresponding to $\bar{v} = 0.215\,\vthe$. In the LOKI
simulations, the velocity $v_c = 6\,\vthe$ (the velocity beyond 
which $\nu_{ei}(v)$ is smoothly ramped down to zero as it approaches $v_{\max}$). 
Note for all cases that, once the oscillatory EPW
  has damped out, an initially small 'collisional' mode with weaker
  damping and zero real frequency survives. This collisional mode, subsequently
  referred to as the entropy mode \cite{Brantov2004}, is not observed for
  zero collisionality. In these figures, presenting waves in the linear
  regime, the decay of the EPW is as expected well fit with an exponential
  decay. 
  The entropy mode appears as well to present a
  constant exponential decay rate. As will be discussed in
  Sec. \ref{sec:entropy}, this constant decay rate of the entropy mode is 
  a consequence of the grid resolution at low velocity. 
  The EPW results from LOKI, requiring
  significantly less fine velocity grids to ensure convergence, are however
  very well resolved. 
 When analyzing the LOKI results, the constant exponential decay
  rate of the entropy mode evolution at later times is extrapolated
  and subtracted from the time traces at earlier times, enabling a
  more accurate determination of the frequency and damping of the EPW. The
time interval over which the EPW frequency and damping are calculated depends 
on the total damping rate; more accuracy is possible at lower rates, with
better than 1\% at $k \lambda_{De} = 0.300$ and $\nuth=0.05$ but less than 3\%
at $k \lambda_{De} = 0.425$ and $\nuth=0.5$.

Real frequencies $\omega_R$ and damping rates $\nu$ of the EPW for $k
\lambda_{De}=0.300$, obtained with both LOKI and the numerical solution of the
system (\ref{Fokker-Planck, cosine component m=0})-(\ref{Poisson Eq. with dfc0
  source}) for different collision frequencies $\nu_{\rm ei, th}/\omega_{pe} =
0$ to $5\cdot 10^{-1}$, are summarized in Fig. \ref{Fig. Coll. effects on
  lin. Landau damp., 2D vs 3D, numax/nuth = 100}.  Numerical solutions to the
linearized Vlasov-Poisson system (\ref{Fokker-Planck, Legendre component
  l=0})-(\ref{Poisson Eq. with df0 source}) for collisions in 3D velocity
space instead of 2D are also shown.  This plot provides an assessment of the
effect on the frequency and damping of the
restriction of the collisional scattering to two velocity dimensions.
Over all the cases considered, the maximum relative difference of the linear
damping rate between the results obtained with the 2D and 3D velocity space
collision operator is only $20\%$ and occurs at high collisionality.
\begin{figure}
  \centering
\includegraphics[width=0.34\textwidth]{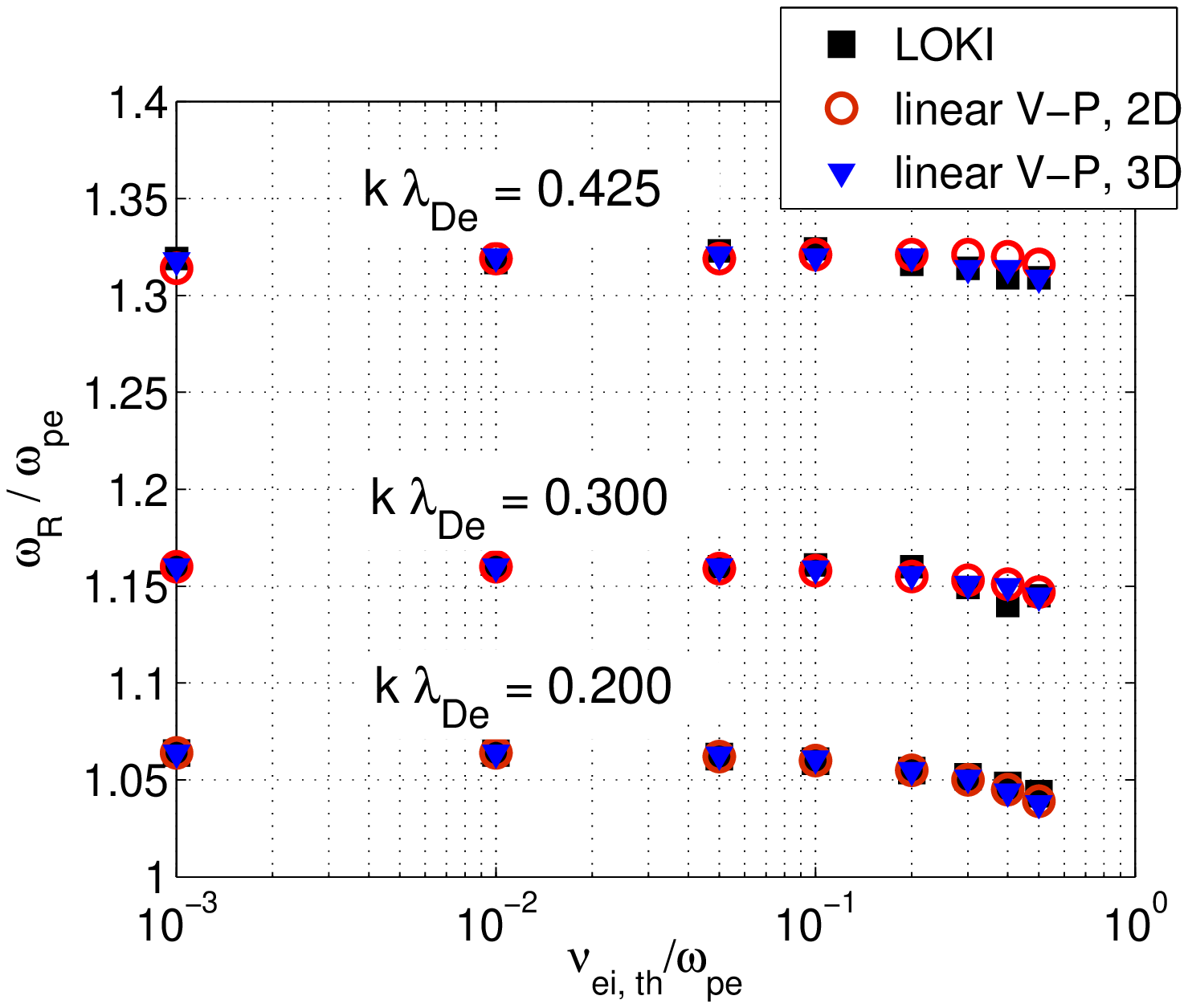}
\includegraphics[width=0.32\textwidth]{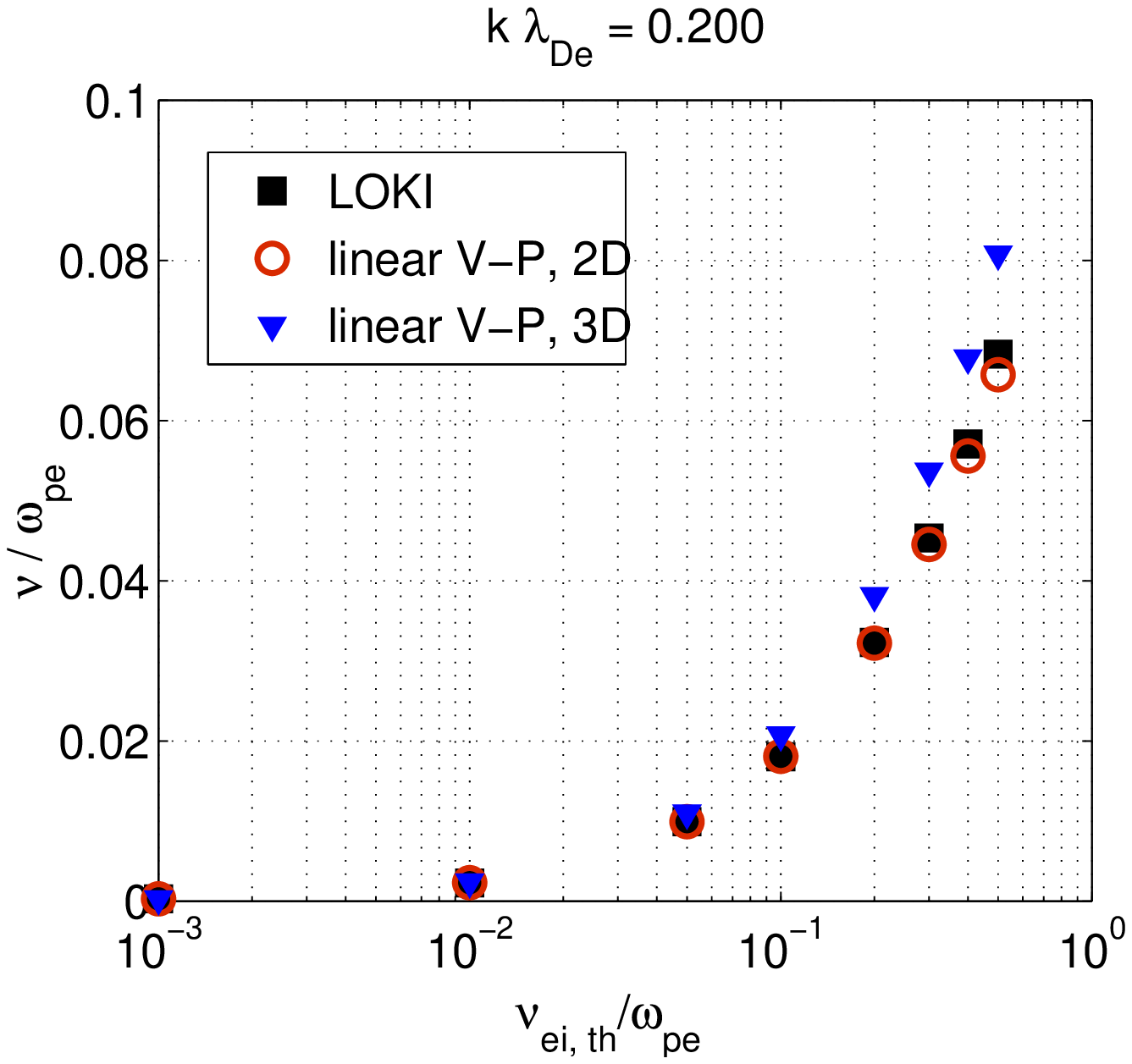}
\includegraphics[width=0.32\textwidth]{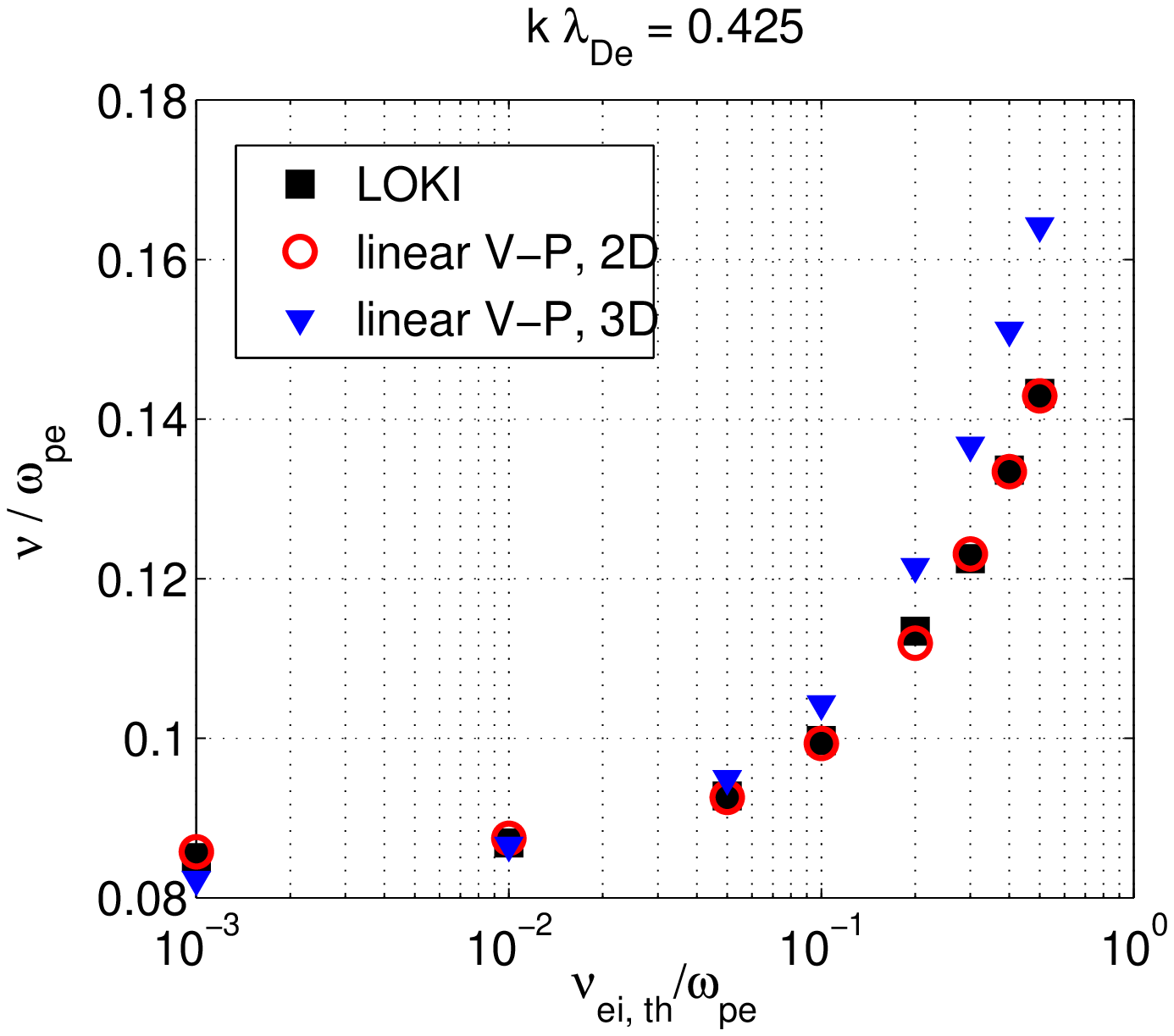}
  \caption[]{\label{Fig. Coll. effects on EPW freq all ks} Effect of
    collisions on real frequency $\omega_R$ and damping rate $\nu$ of an
    EPW. Plotted are results from LOKI simulations (black square markers) as
    well as from the linearized Vlasov-Poisson system of equations with
    collisions in both 2D and 3D velocity space (red circles and blue
    triangles respectively).  In (a), the frequency dependence for $k
    \lambda_{De} = 0.200, 0.300, \rm and\, 0.425$ is shown. In (b) and (c),
    the damping rate for $k\lambda_{De} = 0.200$ and $k\lambda_{De} = 0.425$
    are shown respectively.}
\end{figure}

The effect of collisions on the frequency and damping of EPWs was also studied
with simulations with $k\lambda_{De} = 0.425$, for which the
collisionless linear Landau damping is larger than at $k \lambda_{De} = 0.300$
(phase velocity is $\vph/\vthe = 3.100$ for $k\lambda_{De}=0.425$ instead of
$\vph/\vthe = 3.867$ for $k\lambda_{De} = 0.300$, see Table \ref{linear mode
  characteristics}), as well as with simulations with $k \lambda_{De} =
0.200$, for which the collisionless linear Landau damping is negligible
(corresponding phase velocity $\vph/\vthe = 5.320$, see Table \ref{linear mode
  characteristics}). Figure \ref{Fig. Coll. effects on EPW freq all ks} shows
the results for the frequency and damping as a function of the collision
rate. In Fig. \ref{Fig. Coll. effects on EPW freq all ks}.a, the EPW frequency
as a function of the collision rate $\nuth$ is shown for all the three cases:
$k \lambda_{De} = 0.200$, $0.300$, and $0.425$.  Both the results from LOKI
and from the linearized set are shown. The EPW damping rates from LOKI and the
linearized set are shown in Fig.~\ref{Fig. Coll. effects on EPW freq all ks}.b
and \ref{Fig. Coll. effects on EPW freq all ks}.c for $k \lambda_{De} = 0.200$
and $0.425$ respectively.
%
%
\subsection{Approximate Analytic Solutions for Collisional Damping of EPWs}
\label{collision damping:theory}
The electron-ion collision rate $\nu_{\rm ei, th}$ is typically much smaller
than the collisionless Landau damping rate $\nu_L$ of EPWs. However, in very
dense, high $Z$, or low temperature plasmas, for which
$\nuth$ approaches $\omega_{pe}$, or for very high phase velocity waves, {\it
  i.e.} such that $v_{\phi} \gg v_{th,e}$ in which case Landau damping is
negligibly small, $\nuth$ can be larger than the Landau damping rate
$\nu_L$. Excitation of very high phase velocity EPWs may occur in Forward
Stimulated Raman Scattering (FSRS) or in two-plasmon decay for which Landau
damping of one of the decay EPWs is negligible. For weak collisions,
Refs. \cite{Auerbach77} and \cite{Callen2014} found that collisions do not
affect the magnitude of Landau damping but do cause irreversible dissipation
on a long time scale; thus plasma wave echos may be suppressed.\cite{Su1968}

Collisional damping of EPWs, resulting from the
loss of momentum of the electrons oscillating in the field of the wave, can be
obtained  from the following set of fluid equations that 
represent the electron dynamics in an EPW:
\begin{gather}
\label{fluid eqns1}
\frac{d_e n_e}{dt} + n_e \nabla \cdot \vec{u}_e = 0, \\
\label{fluid eqns2}
m_e n_e \frac{d_e \vec{u}_e}{dt} = e\,n_e \nabla \phi -\nabla p_e -m_e n_e \nu_e^D \left (\vec{u}_e - \vec{u}_i \right ), \\
\frac{d_e }{dt} \left (\frac{p_e}{n_e^{\gamma_e}}\right) = 0,  \\
\label{fluid eqns4}
-\nabla^2 \phi = -4\pi e \left ( n_e -\sum_i N_i Z_i \right ).
\end{gather}
Note in particular the drag term on the right hand side of Eq. (\ref{fluid
  eqns2}), resulting from momentum exchange due to electron-ion pitch angle
scattering.\cite{Braginskii1965} In the system (\ref{fluid eqns1})-(\ref{fluid
  eqns4}), the operator $d_e/dt = \partial/\partial t + \vec{u}_e \cdot
\nabla$ stands for the convective derivative along the mean electron
velocity flow $\vec{u}_e$. The variables $n_e$ and $p_e$ respectively stand
for the density and  pressure of the electrons. The sum on the
right hand side of Eq. (\ref{fluid eqns4}) is again over all ion species $i$,
assumed fixed, with uniform density $N_i$ and charge $Z_ie$, with overall
charge neutrality assumed, $N_e = \sum_i N_i Z_i$, $N_e$ being the average
electron density. Fixed ions imply that their average velocity
$\vec{u}_i=0$. The parameter $\gamma_e = (d+2)/d$ stands for the adiabatic
index, $d$ being the effective dimensionality of velocity space for the
equation of state. Finally, the friction coefficient is denoted $\nu_e^D$. For
three-dimensional pitch-angle scattering, $D=3$, and the friction rate
$\nu_e^{D=3} = \nu_{ei}^{\rm brag}=1/\tau_{e}^{\rm brag}= [4/(3 \sqrt{2
    \pi})]\,\nu_{ei,th}\simeq 0.532\,\nu_{ei,th}$, where $\nu_{ei}^{\rm brag}$
is the electron-ion collision frequency as defined by
Braginski.\cite{Braginskii1965} For two-dimensional scattering, $D=2$, and
$\nu_e^{D=2} = \sqrt{\pi/8}\,\nu_{ei,th}\simeq 0.627\,\nu_{ei,th}$. 
  These friction coefficients $\nu_e^D$ are derived
by evaluating the drag $\vec{R}_{ei} = -m_en_e\nu_e\vec{u}_e = -m_e \int
d^Dv\,\vec{v}\,C_{ei} f_{\rm SM}$ on a linearized shifted Maxwellian electron
distribution $f_{SM} = f_M (1+\vec{v}\cdot\vec{u}_e/\vthe^2)$, with $f_M$
given by Eq. (\ref{normalized Maxwellian in D dimensions}) and the collision
operator $C_{ei}$ by Eq. (\ref{Lorentz coll. op.}). Note that 
$\nu_e^{D=2} > \nu_e^{D=3}$, and, thus, the fluid equations find the EPW collisional 
damping is stronger in 2D than in 3D. Yet, the effect of
collisions on the damping rate observed in the simulations is stronger for
  $D=3$ than for $D=2$. In Appendix \ref{sec:diffusivedamping}, we will show that
the fluid equation damping rate is about twice the correct value although the scaling
with collision frequency is correct. The problem arises in the fact the the drag term 
is strongly weighted by velocities, $ v \sim 0$, whereas, done correctly, velocities
$ \sim 0.2 < v < 1 $ contribute the most. The weighting by higher velocity is stronger in 3D.

The system (\ref{fluid eqns1})-(\ref{fluid eqns4}) can then be linearized with
respect to small fluctuations $\delta n_e$, $\delta\vec{u}_e$ and $\delta p_e$
around the corresponding uniform background electron quantities, {\it i.e.}
density $N_e$, zero background velocity $\vec{U}_e=0$, and background pressure
$P_e = N_eT_e$ respectively, where $T_e$ is the background
temperature. Assuming plane wave fluctuations with frequency $\omega$ and
wavenumber $k$, one obtains a dispersion relation for the real
frequency $\omega_R$ of EPWs given by the Bohm-Gross relation
\begin{equation}
\label{Bohm-Gross disp. rel.}
\omega_R = \omega_{\rm BG} = \omega_{pe} \sqrt{1+\gamma_e (k\lambda_{De})^2}
\end{equation}
and a corresponding damping rate 
\begin{equation}
\label{collisional damping}
\nu_{\rm coll} = (\omega_{pe}/\omega_R)(\nu_{e}^D/2)\simeq \nu_{e}^D/2,
\end{equation}
provided $k\lambda_{De} \ll 1$ and $\nu_e^D \ll \omega_{pe}$. For weakly
collisional plasma, {\it i.e.} $k \lambda_{ei} \gg 1$, with $\lambda_{ei} =
\vthe/\nuth$ the thermal electron-ion mean free path, the effective
dimensionality for estimating $\gamma_e$ is $d=1$, so that
$\gamma_e =3$. For a strongly collisional plasma, {\it i.e.} $k \lambda_{ei}
\ll 1$,  $d=D$, so that $\gamma_e = 5/3$ for
three-dimensional velocity space $D=3$, while $\gamma_e = 2$ for $D=2$.  This
strongly collisional fluid equation result for the frequency and damping rate
in the case $D=3$ is recovered in Appendix\,\ref{diffusion 3D:theory} by
solving a dispersion relation obtained from the $l=0,1$ equations in the
system of Eqs. (\ref{Fokker-Planck, Legendre component l=0})-(\ref{Poisson
  Eq. with df0 source}) in the limit that $k \lambda_{De} \ll 1 $ and $\nuth
\ll \omega_{pe}$. The corresponding two-dimensional ($D=2$) fluid result for
the collisional damping can be recovered as well from the $m=0,1$
equations in the system (\ref{Fokker-Planck, cosine component
  m=0})-(\ref{Poisson Eq. with dfc0 source}), as shown in
Appendix\,\ref{diffusion 2D:theory}. The solution to the dispersion that agrees
with the fluid dispersion is obtained by Taylor expanding
the integrand in the presumably small parameter $\nuth/\omega_{pe}$. 
However, this damping rate is about twice the Vlasov simulation damping rate 
attributable to the direct effect of collisions shown in Fig. \ref{Collisional EPW damping}.
In Appendix \,\ref{diffusion 2D:theory}, we numerically obtain the correct solution 
to the dispersion relation and show that it agrees very well with the kinetic simulations.

%
%
\subsection{Analysis of results}
\label{interpretation}
\begin{figure}
\begin{center}
\includegraphics[width=2.5in]{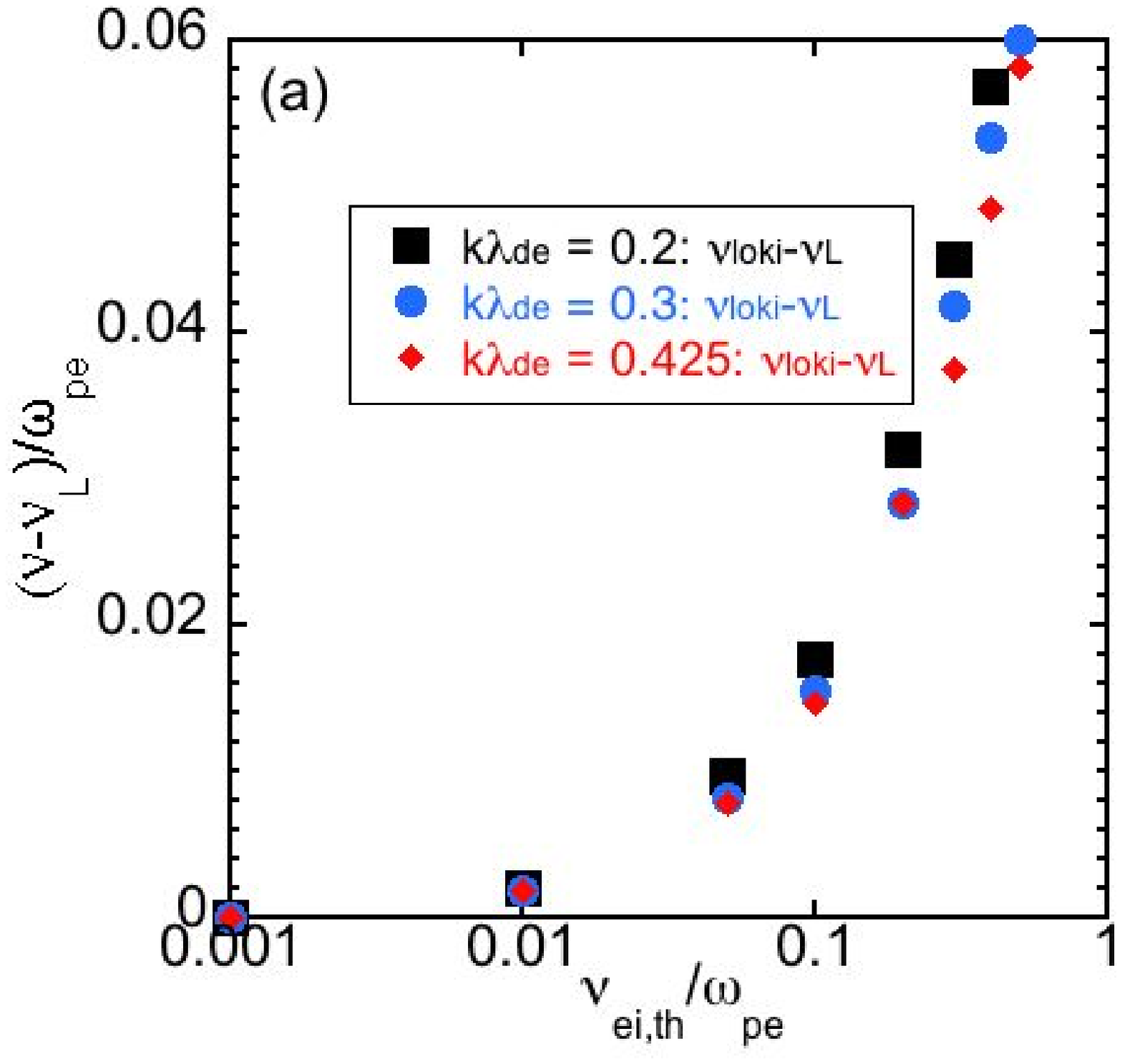}
\includegraphics[width=2.5in]{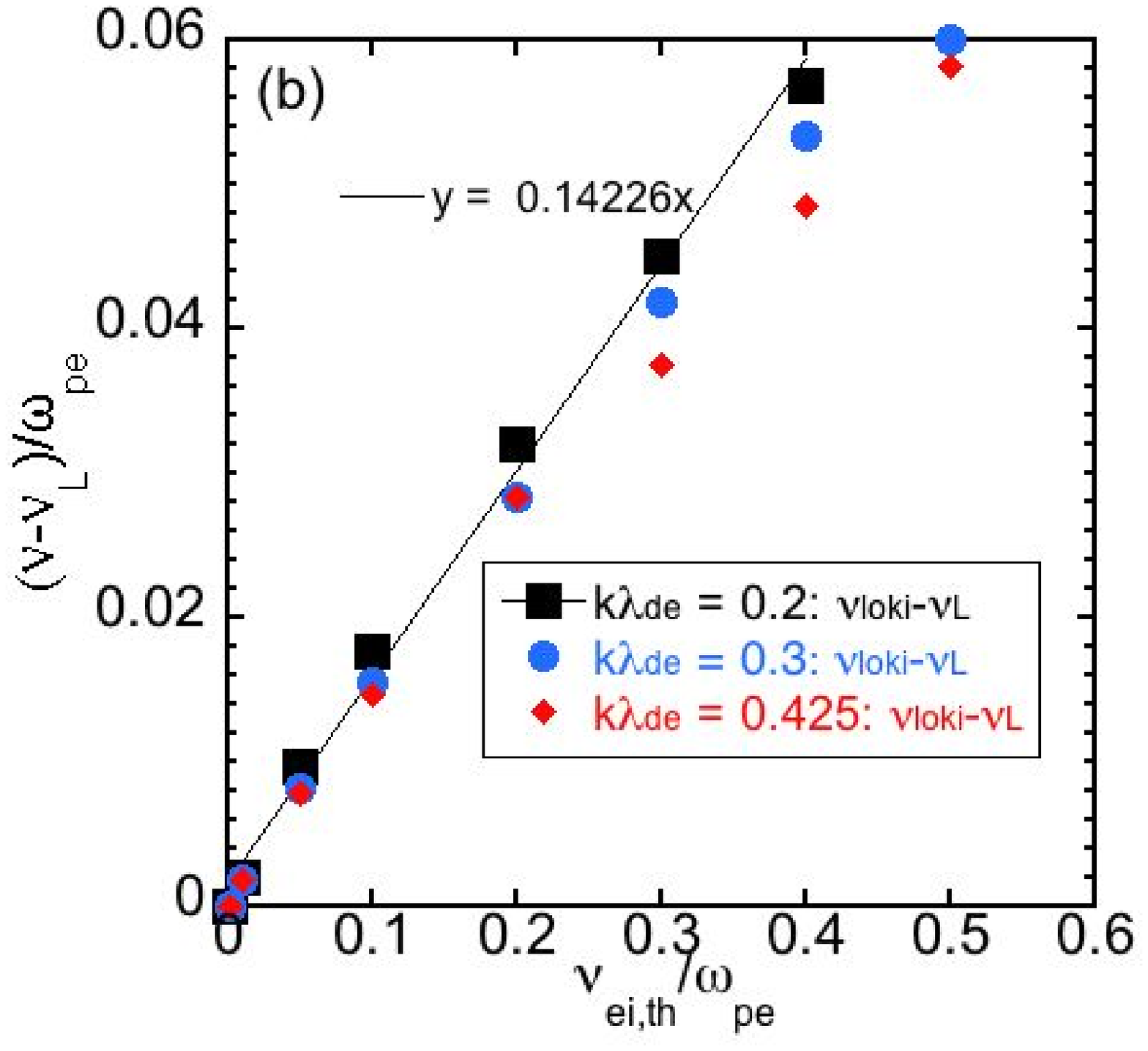}
\caption{\label{Collisional EPW damping} The collisional component $\nu-\nu_L$
  of the EPW damping rate obtained from the LOKI simulations is shown as a
  function of $\nu_{ei,th}/\omega_{pe}$ for the cases $k \lambda_{De} =
  0.200$,$k \lambda_{De} = 0.300$, and $k \lambda_{De} = 0.425$ on (a) a
  logarithmic scale and (b) a linear scale for the collisionality. To obtain
  the collisional damping component, the calculated value
  $\nu_L$ of the collisionless Landau damping (see Table \ref{linear mode
  characteristics}) was subtracted from the total
  damping obtained from the simulations. The collisional component for $k
  \lambda_{De} = 0.200$ follows a linear relation
  wrt. $\nu_{ei,th}/\omega_{pe}$.  The results for the higher $k \lambda_{De}$
  values fall below this line as expected from Eq.~(\ref{collisional damping}).
}
\end{center}
\end{figure}

In Figure \ref{Collisional EPW damping}, the Vlasov simulation results for the
collisional part of the damping rate are shown as a function of
$\nu_{ei,th}/\omega_{pe}$ for the different values of $k
\lambda_{De}$. The collisional component of the damping rate is estimated by
subtracting the collisionless Landau damping rate $\nu_L$ (found analytically,
see Table \ref{linear mode  characteristics}, or equivalently taken from LOKI simulation at small $\nuth$) 
from the total damping rate (shown in
Figs. \ref{Fig. Coll. effects on lin. Landau damp., 2D vs 3D, numax/nuth =
  100} and \ref{Fig. Coll. effects on EPW freq all ks}) obtained from the
simulations. 
Figure \ref{Collisional EPW damping}.a shows no
discernable $k \lambda_{De}$ dependence of the collisional damping component
for $\nu_{ei,th}/\omega_{pe} < 0.1$. For $\nu_{ei,th}/\omega_{pe} > 0.1$, the
EPWs with larger $k \lambda_{De}$ have a lower collisional damping rate
component than the higher phase velocity waves as expected from the factor
$ \omega_{pe}/\omega_R $ in the fluid 
equation result given by Eq. (\ref{collisional damping}).

Note that the collisionless phase velocities are
$\vph/\vthe = 5.320$, $3.867$ and $3.100$ for $k \lambda_{De}= 0.200$, $0.300$
and $0.425$ respectively (see again Table \ref{linear mode characteristics}).
Figure \ref{Fig. Coll. effects on EPW freq all ks}.a shows that the real
frequency $\omega_R$ of the EPWs decreases -- and thus the phase velocity
decreases as well -- with increasing $\nuth$. The Landau damping estimated
with the frequency $\omega_R$ modified by finite collisionality should thus
increase with $\nuth$ and consequently lower the collisional component to the
damping. 

\begin{figure}
\begin{center}
\includegraphics[width=5in]{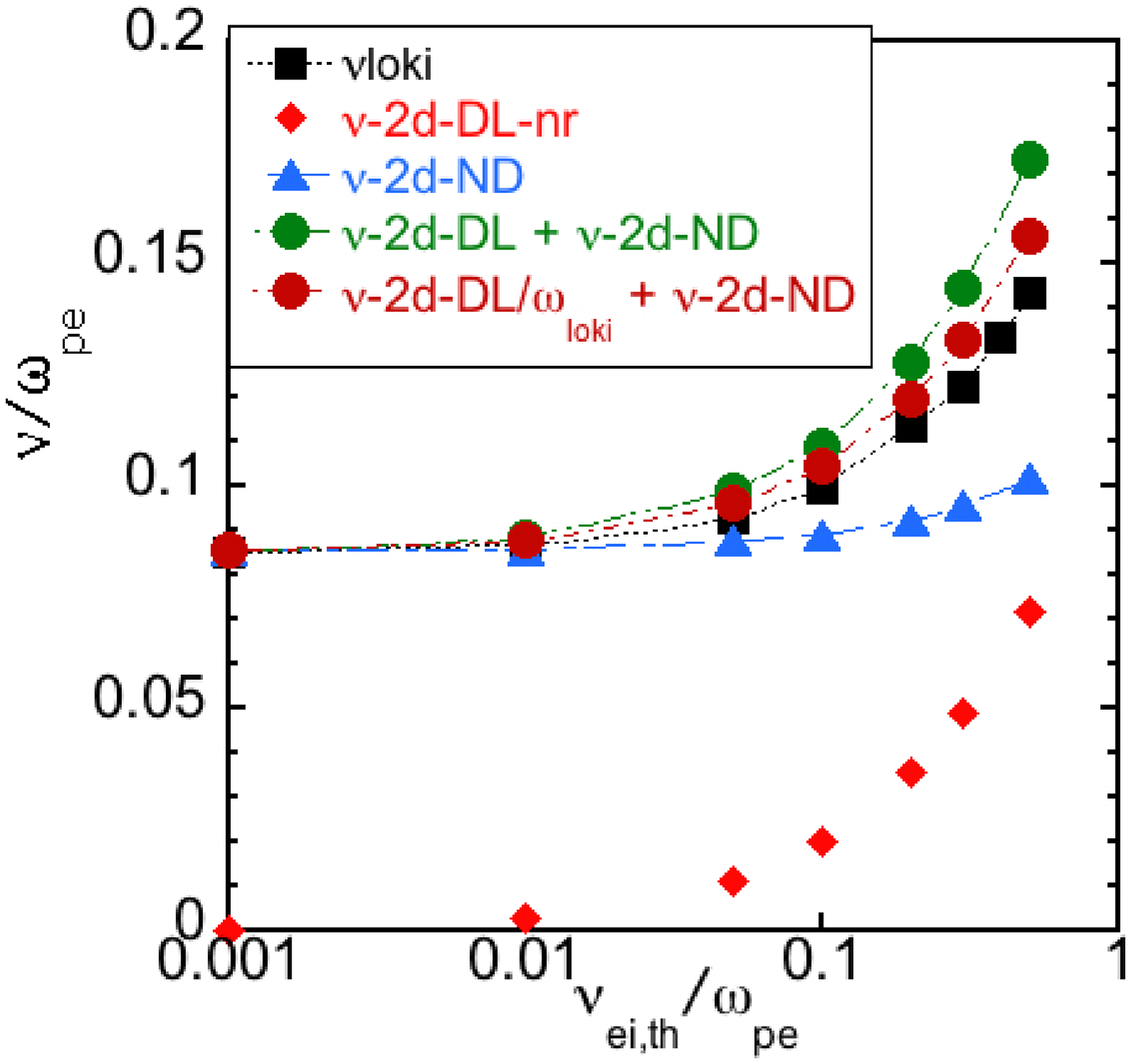}
\caption{\label{limits:EPW damping} Damping rates of EPWs for $k \lambda_{De}
  = 0.425$. Shown are the results from LOKI simulations (black squares) and
  from different limits of the linearized Vlasov-Poisson system (details given
  in text): collisions but without thermal and resonant effects (red
  diamonds); setting the collisional term to zero in the equation for $\delta
  f_{c_1} $ (blue triangles); the sum of red diamonds and blue triangles
  (green circles); and the sum of the blue triangles and red diamonds, the
  latter corrected by the factor $\omega_{pe}/\omega_{R, {\rm LOKI}}$, with
  $\omega_{R, {\rm LOKI}}$ the real frequency obtained from the corresponding
  LOKI simulations (red circles). 
}
\end{center}
\end{figure}

In the following, we consider the effect of arbitrarily setting some terms to
zero in the linearized set of equations (\ref{Fokker-Planck, cosine component
  m=0})-(\ref{Poisson Eq. with dfc0 source}) for the Vlasov-Poisson system
with collisions in two-dimensional velocity space. For this study the
particular case $k\lambda_{De} = 0.425$ is considered. In one limit, the
effect of only collisions, neglecting thermal corrections to the frequency as
well as any wave-particle resonance effects (in particular Landau damping), is
studied by removing the coupling of $\delta f_{c_1}$ to $\delta f_{c_0}$ 
and to $\delta f_{c_2}$ but keeping the coupling 
of $\delta f_{c_0}$ to $\delta f_{c_1}$. That is,  Eq. (\ref{Fokker-Planck, cosine component m=0}) is left unchanged and Eq. (\ref{Fokker-Planck, cosine component m=1}) is replaced by Eq. (\ref{NoLD}):
\begin{eqnarray}
\label{NoLD}
  \frac{\partial\, \delta\!f_{c,1}}{\partial t} 
  - ikv\,\phi\,f_M
  & =
  - \nu_{\rm ei}(v)\,\delta\!f_{c,1}.
\end{eqnarray}
Keeping the coupling of $\delta f_{c_1}$ to $\delta f_{c_0}$ would introduce an
unphysical resonance at $\omega = k v/\sqrt{3} $ in 3D or $\omega = k
v/\sqrt{2} $ in 2D as occurs in Eq. (\ref{3D resonant denominator}) and Eq. (\ref{2D chie})
for 3D and 2D respectively.  
The results of that approximation yield the drag-limit damping rate $\nu_{2dDL}$, 
shown by the red
diamonds in Fig. \ref{limits:EPW damping} and are about 1/2 the
imaginary part of Eq. (\ref{2D collisonal damping}) as explained in Appendix \ref{sec:diffusivedamping}.

In a complementary limit, all terms and equations of the system
(\ref{Fokker-Planck, cosine component m=0})-(\ref{Poisson Eq. with dfc0
  source}) are kept except for the collision term on the right hand side of
the evolution equation for $\delta f_{c_1}$, effectively eliminating drag but
keeping Landau damping. That is,  Eq. (\ref{Fokker-Planck, cosine component m=0}) is left unchanged and Eq. (\ref{Fokker-Planck, cosine component m=1}) 
is replaced by Eq. (\ref{NDiff}): 
\begin{eqnarray}
\label{NDiff}
  \frac{\partial\, \delta\!f_{c,1}}{\partial t} 
  + ikv\,
  \left(
  \delta\!f_{c,0} + \frac{\delta\!f_{c,2}}{2}
  \right)
  - ikv\,\phi\,f_M
  & = 0.
\end{eqnarray}
The results of that approximation yield the no-drag damping rate $\nu_{2dND}$,  
shown by the blue triangles in
Fig. \ref{limits:EPW damping}. This latter approximation is thus meant to
include Landau damping and the effect of collisions on Landau damping but not
collisional damping resulting from drag on the non-resonant electrons, at the
origin of the damping derived in Sec. \ref{collision damping:theory}.

The sum of the damping from the two above-mentioned approximations yields the
green circles in Fig. \ref{limits:EPW damping}, which define a set of damping
rates which are close but somewhat larger than the values from the LOKI
simulations shown by the black squares in this same figure. The numerical
solution of the full linearized set (\ref{Fokker-Planck, cosine component
  m=0})-(\ref{Poisson Eq. with dfc0 source}) agrees very well with LOKI
results as was shown in Figs.~\ref{Fig. Coll. effects on lin. Landau damp., 2D
  vs 3D, numax/nuth = 100} and \ref{Fig. Coll. effects on EPW freq all ks}.
Self-consistency is one difficulty in making the comparison in
Fig. \ref{limits:EPW damping} between the green data points and the simulation
results. As already discussed above, Landau damping depends on the phase
velocity $\vph =\omega_R/k$ which is determined by the solution to the
linearized set that has been modified by dropping terms. Analysis of results
for the real frequency $\omega_R$ without the collision term in the equation
for $ \delta f_{c_1}$ shows that it is nearly the same as the LOKI frequency,
$\omega_{R, {\rm LOKI}}$, so the blue triangles are unaffected by a $\vph$
difference. The real frequency $\omega_R$ in the purely collisional case is
however quite different, essentially $\omega_R = \omega_{pe}$, because there
are no thermal contributions to the dispersion. 
Using the relation for the collisional damping from Eq. (\ref{2D collisonal damping}),
$\nu_{\rm coll} = \sqrt{\pi/2}{\nu_{ei,th}}/{4\omega_R}$, we reduced the
collision damping contribution to the total damping by multiplication with the
factor $\omega_{pe}/\omega_{R, {\rm LOKI}}$, which brings the results 
[red open circles in Fig. \ref{limits:EPW damping}]
close to the LOKI results. Here $\omega_{R, {\rm LOKI}} $ is the frequency in the corresponding LOKI simulation. 
The ansatz that collisions would reduce Landau damping, that is, that the blue
triangles would define a decreasing sequence of points in Fig. \ref{limits:EPW
  damping}, is not borne out by this analysis perhaps because the phase
velocity in the linearized system decreases as the collision rate increases.
%
%
%
\subsection{Entropy mode}
\label{sec:entropy}

The entropy mode can be conveniently modeled in the so-called diffusive
limit. The focus here is on the case of electron-ion pitch angle scattering in
two-dimensional velocity space, $D=2$, relevant to the results obtained by the
LOKI code. The starting point are Eqs. (\ref{Fokker-Planck, cosine component
  m=0}) and (\ref{Fokker-Planck, cosine component m=1}) for the two
coefficients $\delta f_{c,0}(v)$ and $\delta f_{c,1}(v)$ of the fluctuating
part of the electron distribution decomposed in polar Fourier modes (here in 
normalized units):
\begin{gather}
  \label{diffusive limit, cosine component m=0}
  \frac{\partial\, \delta\!f_{c,0}}{\partial t}
  + \frac{ikv}{2}\delta f_{c,1}
  =0,\\
  \label{diffusive limit, cosine component m=1}
  ikv\,(\delta f_{c,0} - \phi\,f_M)
  =
  - \nu_{\rm ei}(v)\,\delta f_{c,1}. 
\end{gather}
In Eq. (\ref{diffusive limit, cosine component m=1}), the term $\partial\,
\delta\!f_{c,1}/\partial t$, related to electron inertia, has been neglected
with respect to the collision term on the right hand side of this equation,
under the assumption that the perturbation varies slowly on the electron-ion
collision time scale $\nuth$. Also, the coupling to $\delta\!f_{c,2}$ has been
neglected in this same equation, which is justified under the further
assumption $k\lambda_{\rm ei}\ll 1$, corresponding to the collisional limit.
The ratio of consecutive coefficients indeed scales as $|\delta f_{c,
  m+1}/\delta f_{c,m}| \sim k\lambda_{\rm ei}$. Eliminating $\delta f_{c,1}$
from the system (\ref{diffusive limit, cosine component m=0}) and
(\ref{diffusive limit, cosine component m=1}) and furthermore making use of
the Poisson equation (\ref{Poisson Eq. with dfc0 source}) leads to an
effective equation for the component $\delta f_{c,0}(v, t)$:
\begin{equation}
  \label{Eq. for df0, diffusive approx., 2D}
  \frac{\partial\, \delta\!f_{c,0}}{\partial t}
  =
  -\frac{(kv)^2}{2\,\nu_{\rm ei}(v)}\delta f_{c,0}
  -
  \pi \frac{v^2 f_M(v)}{\nu_{\rm ei}(v)} 
  \int_0^{+\infty} \hspace{-0.4cm}v\,dv\, \delta\!f_{c,0}\quad,
\end{equation}
referred to as the diffusive approximation.\cite{Albritton1983,Epperlein94}

The evolution equation (\ref{Eq. for df0, diffusive approx., 2D}) has a
spectrum of eigenmodes with fixed damping rate $\nu\in\mathbb{R}_+$:
\begin{equation}
  \label{eigenmode, diffusive approx., 2D}
  \delta f_\nu(v, t)
  =
  \delta \hat{f}_\nu(v)e^{-\nu t}
  =
  \left[
    \pi\frac{v^2f_M(v)}{\nu_{\rm ei}(v)}
    \frac{\mathcal{P}}{\nu - (kv)^2/[2\nu_{\rm ei}(v)]}
    +
    \frac{A}{v_0}\,
    \delta(v-v_0)
    \right]
  e^{-\nu t}.
\end{equation}
In Eq. (\ref{eigenmode, diffusive approx., 2D}), the velocity
$v_0\in\mathbb{R}_+$ is defined such that $\nu = (kv_0)^2/[2\nu_{\rm
    ei}(v_0)]$, or equivalently $v_0/\vthe =
[2\,\nu\,\nuth/(k\vth)^2]^{1/5}$. The symbol $\mathcal{P}$ appearing in the
first term on the right hand side of Eq. (\ref{eigenmode, diffusive approx.,
  2D}) indicates that the resonance at $v_0$ in the denominator is to be
handled in the sense of the Cauchy principal value when carrying out velocity
integrals of this term. The coefficient $A$ multiplying the Dirac function
centered at $v_0$ in the second term on the right hand side of
Eq. (\ref{eigenmode, diffusive approx., 2D}) is given by
\begin{equation}
  \label{coeffA for eigenmode, diffusive approx., 2D}
  A 
  = 
  1
  -
  \pi\,\mathcal{P}\hspace{-4.mm}\int_0^\infty \hspace{-3.mm} dv
  \frac{v^3f_M(v)}{\nu_{\rm ei}(v)}
  \frac{1}{\nu - (kv)^2/[2\nu_{\rm ei}(v)]}.
\end{equation}
One can easily check by substitution that Eqs. (\ref{eigenmode, diffusive
  approx., 2D}) and (\ref{coeffA for eigenmode, diffusive approx., 2D})
provide a solution to (\ref{Eq. for df0, diffusive approx., 2D}) for any
$\nu\in\mathbb{R}_+$. The spectrum of the diffusive equation is thus purely
real, positive and continuous. The associated eigenmodes of the form
(\ref{eigenmode, diffusive approx., 2D}) are clearly singular at $v=v_0$. One
may note the analogy between these eigenmodes to the diffusive
Eq. (\ref{Eq. for df0, diffusive approx., 2D}) and the singular Van Kampen
eigenmodes to the collisionless Vlasov-Poisson system,\cite{VanKampen1955} the
corresponding spectra in this latter case being also continuous, however
purely imaginary (reflecting undamped modes). The fact that the eigenmodes of
Eq. (\ref{Eq. for df0, diffusive approx., 2D}) are singular and the associated
spectrum is continuous results from the fact that only pitch angle scattering
collisions have been considered here. Accounting for thermalization of the
electrons through self-collisions would lead to a related diffusive term
involving a second order derivative in Eq. (\ref{Eq. for df0, diffusive
  approx., 2D}). The nature of the equation would consequently be modified,
leading to a discrete spectrum and associated non-singular eigenmodes.

For a smooth initial perturbation $\delta f_{c,0}(v, t=0)$, such as the
sinusoidal density perturbation considered for the EPW simulations and given
by (\ref{initial condition on dfc0 and dfs0}), the evolution predicted by the
diffusive approximation (\ref{Eq. for df0, diffusive approx., 2D}) will be an
infinite superposition over the continuous spectra of eigenmodes of the form
(\ref{eigenmode, diffusive approx., 2D}) with different values of $\nu$:
$\delta f_{c,0}(v, t) = \int_0^\infty d\nu \,C(\nu)\,\delta
\hat{f}_\nu(v)\,e^{-\nu t}$, with $C(\nu)$ the coefficient function of the
decomposition. The evolution of such a solution will thus not present a single
exponential decay, it being a continuous superposition of modes with different
damping rates $\nu$.

\begin{figure}
\begin{center}
\includegraphics[width=3.5in]{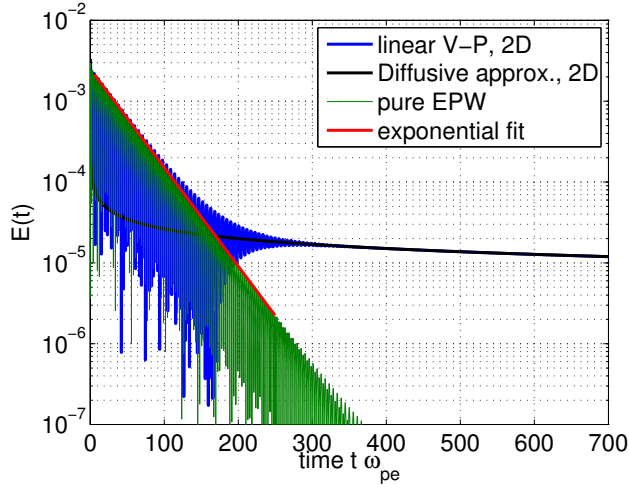}
\caption{\label{Fig. Subtracting entropy mode} Solutions to both the
  linearized Vlasov-Poisson system with collisions (\ref{Fokker-Planck, cosine
    component m=0})-(\ref{Poisson Eq. with dfc0 source}) (blue line) and the
  diffusive approximation equation (\ref{Eq. for df0, diffusive approx., 2D})
  (black) for the same initial sinusoidal density perturbation in case of 2D
  velocity scattering. Plotted is the evolution of the perturbation amplitude
  as a function of time in lin-log scale. Subtracting the two solutions
  solution (black from blue curve) provides the time evolution of the pure EPW
  (green), presenting an exponential damping over a much longer time. A fit
  (red) to the envelope of this latter curve provides a very accurate estimate
  for the decay rate of the EPW. In this example, $k\lambda_{De} = 0.300$,
  $\nu_{\rm ei, th}/\omega_{pe} = 1\cdot 10^{-1}$ and $\nu_{\rm
    ei,\max}/\nu_{\rm ei, th} = 100$.}
\end{center}
\end{figure}

A typical solution to the diffusive approximation equation (\ref{Eq. for df0,
  diffusive approx., 2D}) for an initial perturbation of the form
(\ref{initial condition on dfc0 and dfs0}) is given in Figure
\ref{Fig. Subtracting entropy mode}, for a perturbation with
$k\lambda_{De} = 0.300$, collisionality $\nu_{\rm ei, th}/\omega_{pe} = 1\cdot
10^{-1}$ and $\nu_{\rm ei,\max}/\nu_{\rm ei, th} = 100$. This numerical result
was obtained with a velocity amplitude grid resolution of $n_v=256$ and
$v_{\rm max}/\vthe = 7$. Plotted is the amplitude of the electrostatic field
as a function of time. Note that the evolution is purely damped, {\it
  i.e.} non-oscillatory, reflecting, as expected, that the diffusive
approximation model does not reproduce the EPW dynamics. Furthermore,
 the perturbation is very strongly damped at early times, due to the decay
of the eigenmode components $C(\nu)\delta f_\nu$ related to high values of
$\nu$. These highly damped eigenmodes are singular at a high velocity value
$v_0/\vthe = [2\,\nu\,\nuth/(k\vth)^2]^{1/5}$, such that corresponding
effective collisionality is low, $\nu_{\rm ei}(v_0) \ll \nu$, and consequently
the effective mean free path is large, {\it i.e.} $k v_0/\nu_{\rm ei}(v_0) \gg
1$. These scalings violate the assumptions made in deriving the diffusive
model. At later times, the remaining perturbation from the diffusive
approximation result decays at a much slower rate, corresponding to the decay
of the eigenmode components $C(\nu)\delta f_\nu$ with low values of
$\nu$. These weakly damped eigenmodes are singular at a low velocity value
$v_0$, with a high corresponding effective collisionality $\nu_{\rm ei}(v_0)
\gg \nu$ and a short effective mean free path, $k v_0/\nu_{\rm ei}(v_0) \ll
1$. These latter scalings are in agreement with the assumptions of the
diffusive model.

Also shown in Fig. \ref{Fig. Subtracting entropy mode} is the evolution of the
linearized Vlasov-Poisson system with collisions given by
Eqs. (\ref{Fokker-Planck, cosine component m=0})-(\ref{Poisson Eq. with dfc0
  source}), with oscillating EPW evolution at early times and purely damped
entropy mode evolution at later times. This numerical result was obtained with
a velocity amplitude grid resolution of $n_v=256$, $v_{\rm max}/\vthe = 7$ and
maximum $M=8$ polar Fourier modes. Note how the later time evolution is
perfectly reproduced by the solution to the diffusive approximation
equation. This agreement at later times is expected, given that, as already
mentioned, it corresponds to the evolution which is within the limits of
validity of the reduced diffusive approximation model.

Subtracting over the full simulation time the entropy mode evolution provided by the diffusive approximation from the
evolution of the full Vlasov-Poisson system with collisions
provides the evolution of the EPW  with a single
exponential decay rate over a very long simulation time, as shown in
Fig. \ref{Fig. Subtracting entropy mode}. Such a subtraction
technique enables 
a very accurate estimate of the real frequency and
damping rate for the EPW and was applied for all results presented in figures
\ref{Fig. Coll. effects on
  lin. Landau damp., 2D vs 3D, numax/nuth = 100} and \ref{Fig. Coll. effects
  on EPW freq all ks}.

\begin{figure}
\begin{center}
\includegraphics[width=3.5in]{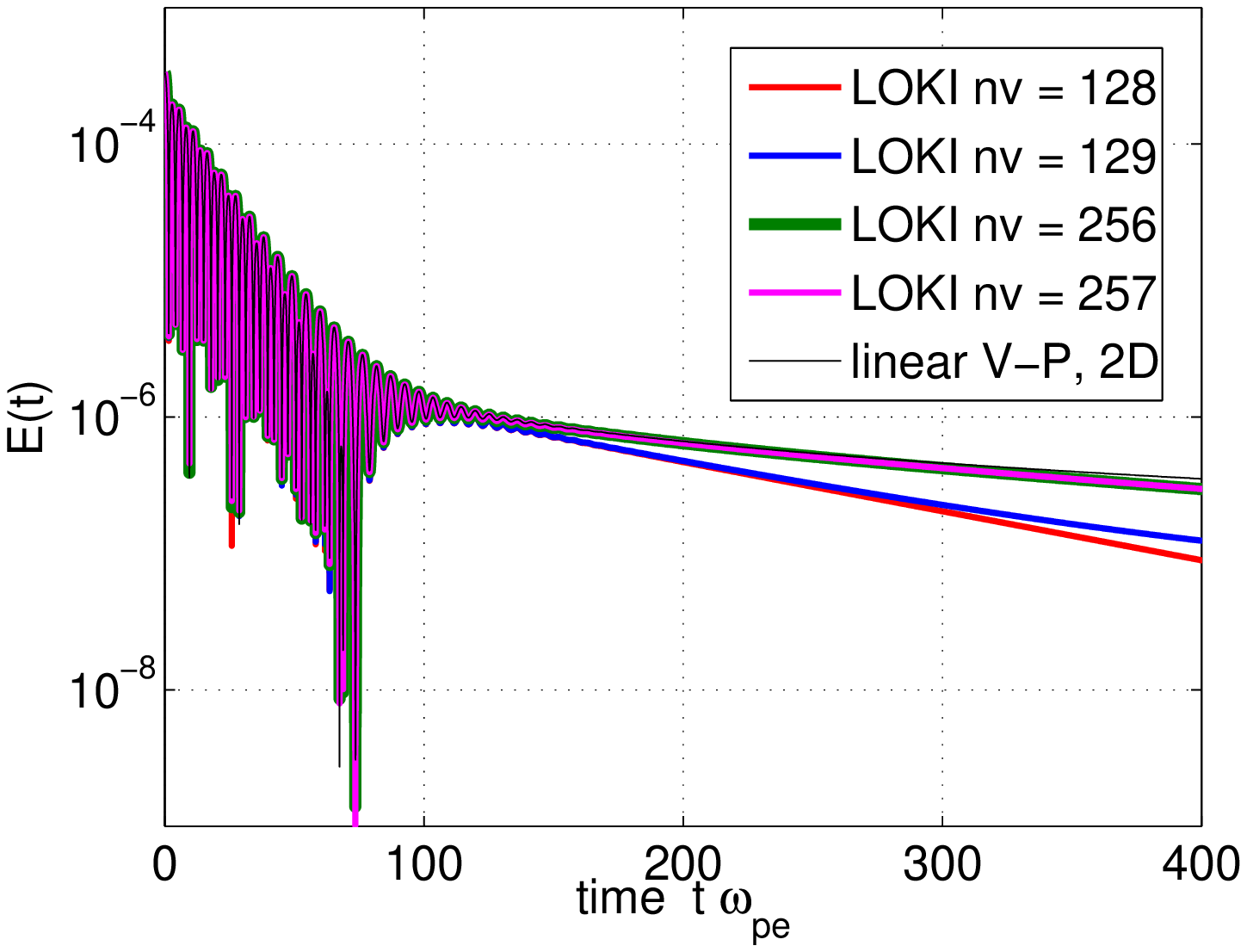}
\caption{\label{Fig. Convergence study of entropy mode} Convergence study in
  case of the entropy mode for LOKI simulations. Shown is the time evolution
  of the mode amplitude for the case $k\lambda_{De} = 0.300$, $\nuth =
  5\cdot 10^{-1}\omega_{pe}$ and $\nu_{\rm ei, \max} = \nuth$, corresponding
  to $\bar{v} = \vthe$. Resolution is simultaneously increased along both
  $v_x$ and $v_y$, with $n_{v_x} = n_{v_y} = 128$, $129$, $256$, and $257$
  grid points and keeping $v_{x, \max} = v_{y, \max} = v_{\max} =7\,\vthe$
  fixed. Also plotted is the corresponding result from the linearized
  Vlasov-Poisson system (\ref{Fokker-Planck, cosine component
    m=0})-(\ref{Poisson Eq. with dfc0 source}) in polar velocity coordinates,
  obtained with the following resolution ensuring convergence: $v_{\rm max} =
  7$, $n_v = 128$ and maximum polar Fourier mode $M=8$.}
\end{center}
\end{figure}

The derivation of the reduced diffusive approximation model
(\ref{Eq. for df0, diffusive approx., 2D}) in case of three-dimensional velocity
scattering, $D=3$, can obviously also be carried out 
starting from Eqs. (\ref{Fokker-Planck, Legendre component
  l=0}) and (\ref{Fokker-Planck, Legendre component l=1}) for the evolution of
the first two components $\delta f_0(v)$ and $\delta f_1(v)$ of the
fluctuating part of the electron distribution decomposed in Legendre
polynomials. 
The eigenmodes 
again define a
purely real, positive, continuous spectrum $\nu\in\mathbb{R}_+$ with
associated singular eigenmodes:
\begin{equation}
  \label{eigenmode, diffusive approx., 3D}
  \delta f_\nu(v, t)
  =
  \delta \hat{f}_\nu(v)e^{-\nu t}
  =
  \left[
    \frac{2\pi}{3}\frac{v^2f_M(v)}{\nu_{\rm ei}(v)}
    \frac{\mathcal{P}}{\nu - (kv)^2/[6\nu_{\rm ei}(v)]}
    +
    \frac{A}{v_0^2}\,
    \delta(v-v_0)
    \right]
  e^{-\nu t},
\end{equation}
the coefficient $A$ being given in this case by
\begin{equation}
  \label{coeffA for eigenmode, diffusive approx., 2D}
  A 
  = 
  1
  -
  \frac{2\pi}{3}\,\mathcal{P}\hspace{-4.mm}\int_0^\infty \hspace{-3.mm} dv
  \frac{v^4f_M(v)}{\nu_{\rm ei}(v)}
  \frac{1}{\nu - (kv)^2/[6\nu_{\rm ei}(v)]}.
\end{equation}
Let us finally address the convergence of the late time entropy mode evolution
in the LOKI simulations. As already mentioned in Sec. \ref{damping:sims} when
commenting on the results plotted in Fig. \ref{Fig. lin. Landau damp.,
  with/without coll.}, the entropy mode evolution is not fully resolved in
most LOKI simulations. This can be better understood based on the above
discussion of the entropy eigenmodes (\ref{eigenmode, diffusive approx., 2D})
derived in the diffusive approximation. It has been shown that the late time
evolution of the Vlasov-Poisson system with collisions is defined by the
slowest decaying entropy eigenmodes which are singular at near zero velocity
$v_0\ll\vthe$. As these eigenmodes are very well represented by the diffusive
approximation, which considers only poloidal Fourier modes $m=0$, $1$, they
are nearly circularly symmetric and  easily resolved when solving the
linearized Vlasov-Poisson system in polar velocity coordinates, as considered
for the system of Eqs. (\ref{Fokker-Planck, cosine component
  m=0})-(\ref{Poisson Eq. with dfc0 source}) ( {\it e.g.}  the green
curve in Fig. \ref{Fig. Subtracting entropy mode}). However, the
Cartesian velocity coordinate system used by LOKI is not appropriate for easily
resolving the singular structure of weakly damped entropy eigenmodes in the
vicinity of zero velocity, as the number of grid points in this region of
velocity space with this representation is very limited for typical
resolutions required for accurately evolving EPWs. Nonetheless, convergence of the LOKI
simulation results with respect to increasing velocity resolution is
illustrated in Figure \ref{Fig. Convergence study of entropy mode}. Plotted is
the perturbation amplitude evolution of an initial sinusoidal density
perturbation with $\delta n/N = 1\cdot 10^{-4}$ (ensuring linear regime),
$k\lambda_{De}=0.300$, $\nuth = 5\cdot 10^{-1}\omega_{pe}$ and $\nu_{\rm
  ei,\max}=\nu_{\rm ei, th}$. The resolution along the $v_x$ and $v_y$
velocity directions were the same. In a series of
computations, an increasing number of velocity grid points $n_v=128$, $129$,
$256$, and $257$ were chosen. In all cases the maximum velocity grid value was
set to $v_{x, \max} = v_{y, \max} = v_{\max} = 7\vthe$. For the lowest
resolution, $n_v=128$, the early time EPW evolution is already fully
converged. The later time entropy mode evolution presents however a constant
exponential decay, illustrating that the simulation was only able to resolve
some of the infinite number of weakly damped entropy eigenmodes located near
zero velocity. As the velocity resolution is increased, additional weakly
damped entropy eigenmodes are resolved by LOKI, and one consequently observes
that the corresponding perturbation amplitude presents a time evolution with
varying decay rate. For $n_v=256$ the LOKI result is thus nearly converged
with the reference result provided by the linearized Vlasov-Poisson system in
polar coordinates given by (\ref{Fokker-Planck, cosine component
  m=0})-(\ref{Poisson Eq. with dfc0 source}) over the  simulation
time $0 < t\omega_{pe} < 400$. Fig. \ref{Fig. Convergence study of
  entropy mode}, shows a significant effect on the entropy mode evolution 
for an odd number of grid points for which a grid point at the critical 
velocity point $(v_x, v_y) = (0,0)$ exists near the location of 
the weakest damped eigenmodes.
Such a grid point at zero velocity is lacking for $n_v$ even.

%
\section{Simulation Results of Nonlinear Landau Damping}
\label{sec:NonlinearEPW}
The results presented in the previous sections addressed the linear regime of
very small amplitude waves for which the linear EPW damping rate $\nu$ is much
faster than the bounce frequency $\omega_B$ of an electron trapped in the
potential well of the wave. The bounce frequency may be estimated with the
deeply trapped estimate $\omega_B/\omega_{pe}=k \lambda_{De} \sqrt{e
  \phi_0/T_e}$, where $\phi_0$ is the finite amplitude of the electrostatic
potential associated to the wave.

Without collisions, a wave initialized with amplitude $\phi_0$ such that
$\omega_B > \nu$ will damp at the linear Landau rate $\nu_L$ only until the
bounce time, $\tau_B \sim \omega_{Be}^{-1}$.\cite{ONeil65,Morales72} After
that, the electrostatic field and bulk electrons exchange energy back and
forth, but without further net transfer to resonant particles and consequently
there is no further Landau damping of the wave.  In
Fig. \ref{Fig. non-lin. Landau damp., with/without coll.}, the time history of
a wave with amplitude $\delta n/N = 0.1 $ in a collisionless plasma shows the
wave reaches such a stationary state, referred to as a BGK
mode.\cite{Bernstein57}
\begin{figure}[p]
  \centering
 \includegraphics[width=0.35\textwidth]{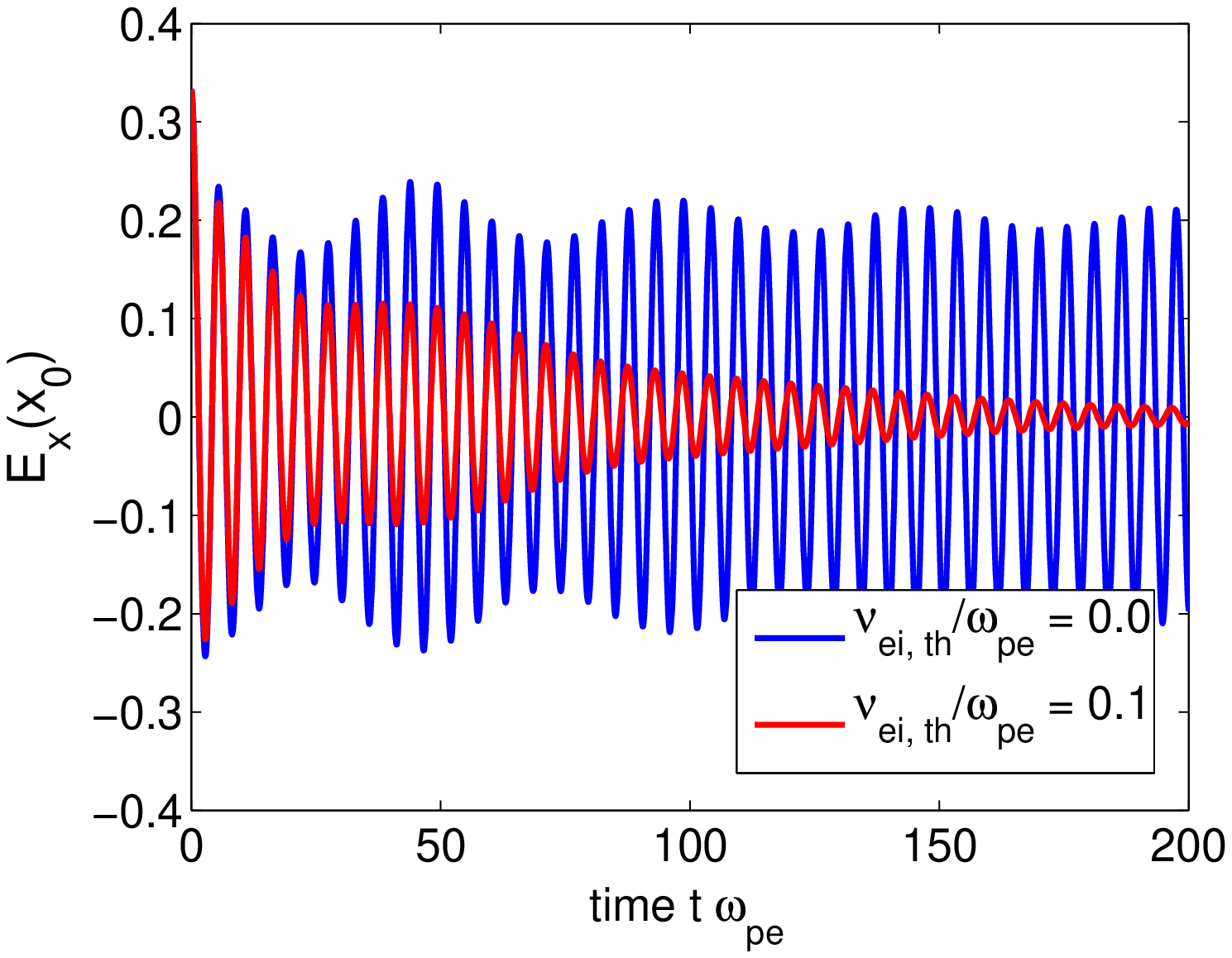}
 \includegraphics[width=0.3\textwidth]{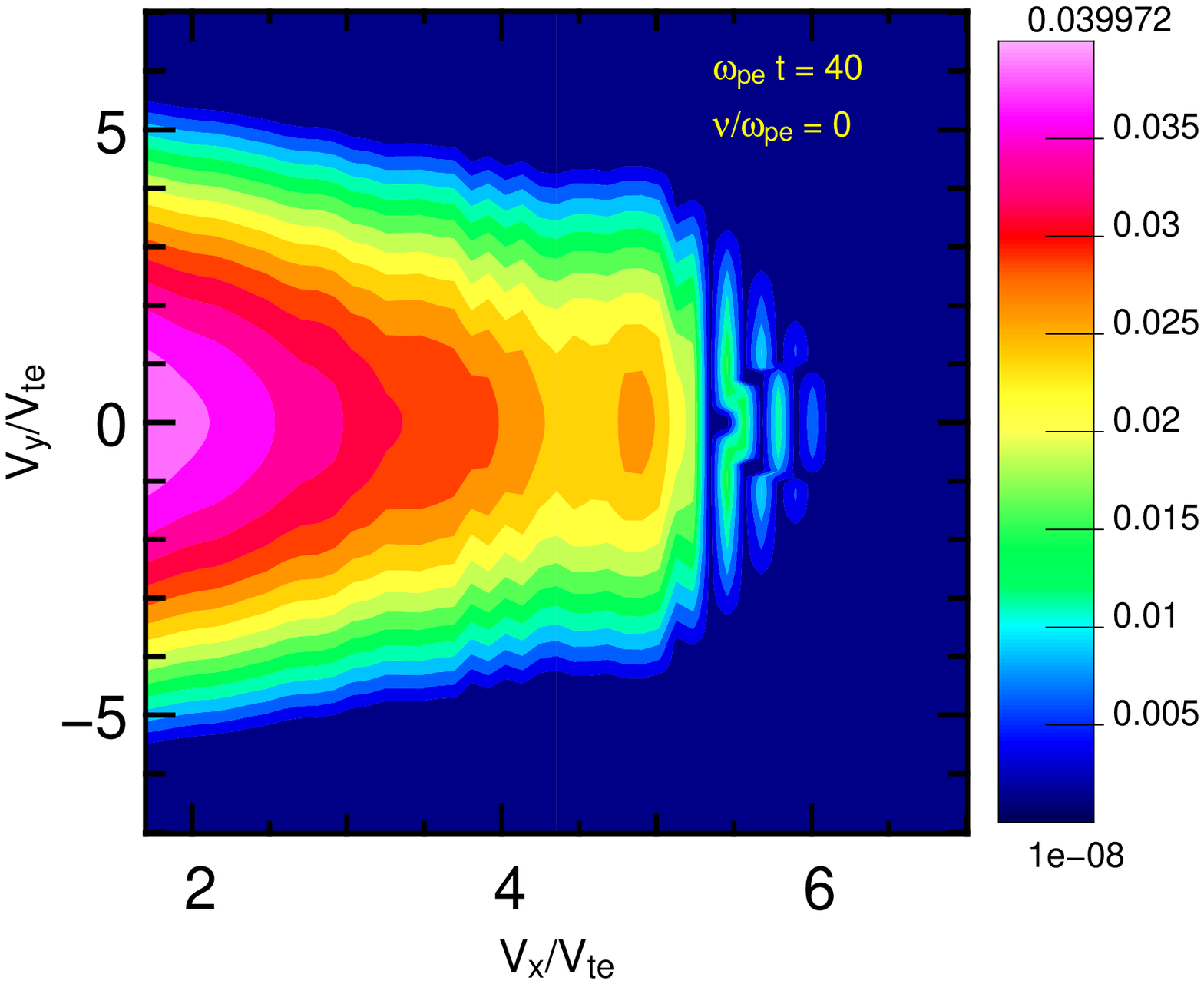}
 \includegraphics[width=0.3\textwidth]{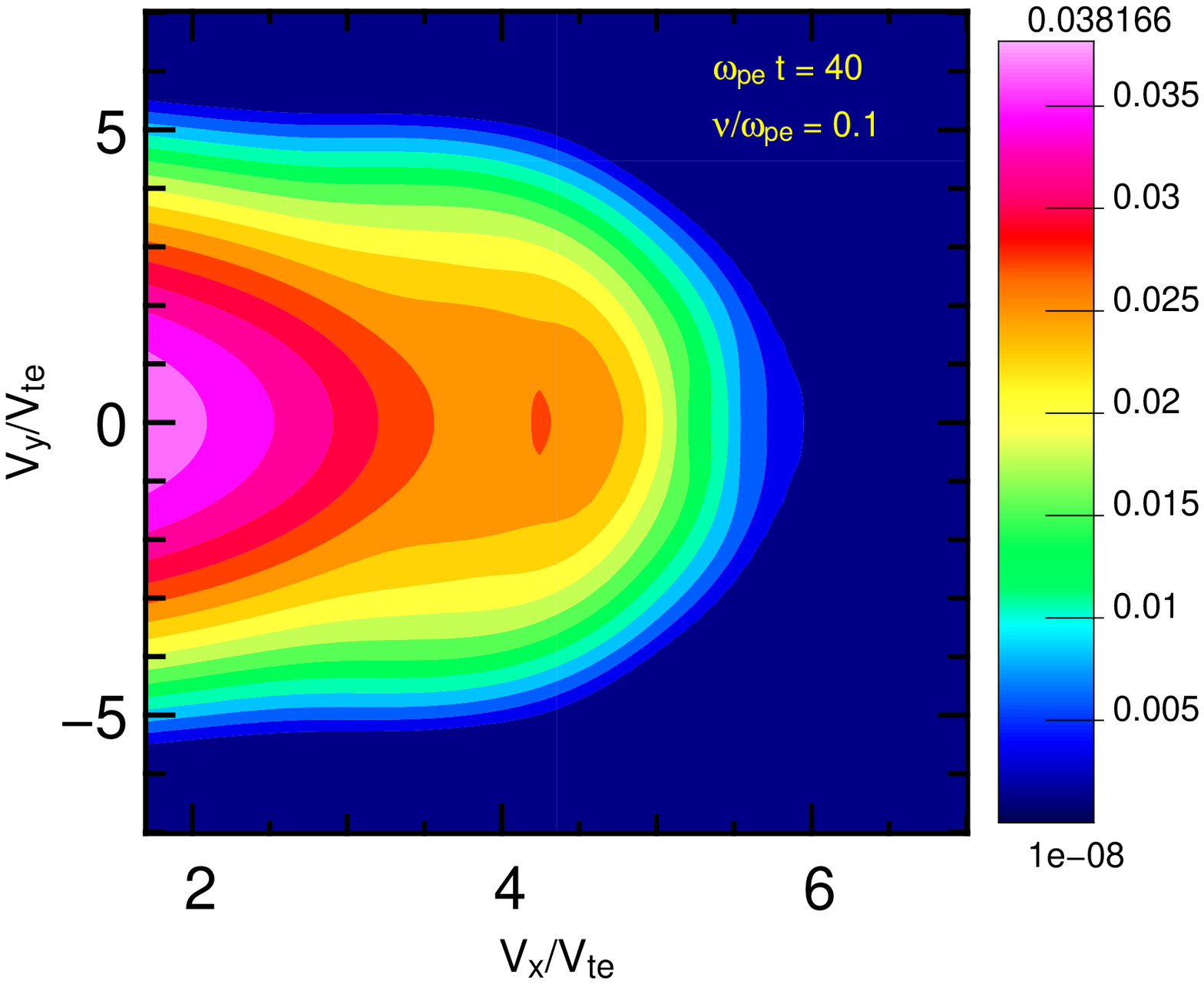}
   \caption[]{\label{Fig. non-lin. Landau damp., with/without coll.}  Results
     from the LOKI code of non-linear Landau damping.  (\textit{left})
     The amplitude of the electrostatic field $E_x$ (longitudinal component) 
     as a function of time $\omega_{pe} t$ 
     for an EPW with $k\lambda_{De} = 0.3$ and
     initial relative amplitude of the density perturbation $\delta n/N =
     0.1$. Results from both a simulation with zero collision rate (blue) and
     finite collision rate $\nu_{\rm ei, th}/\omega_{pe} = 0.1$ (red) are
     shown. The electron distribution function as a function 
     of velocity $v_x$ and $v_y$ averaged over $x$ in the resonant region
     is shown at $\omega_{pe} t = 40$  (\textit{middle}) for the collisionless 
     case and  (\textit{right}) the collisional case with $\nu_{ei,th} = 0.1$.}
\end{figure}
With strong collisions (in this case $\nu_{ei,th}/\omega_{pe}=0.1$), 
the wave continues to damp after a bounce time.
Results of collisional effects on non-linear Landau damping obtained with the
LOKI code are provided in Fig. \ref{Fig. non-lin. Landau damp., with/without
  coll.}, where the time trace of the longitudinal field $E_x$ at the position
$x$ of the anti-node of the standing wave is shown for an initial relative
amplitude of the density perturbation $\delta n/N = 0.1$, both in the case of
zero collision rate, and finite collision rate $\nu_{\rm ei, th}/\omega_{pe} =
0.1$. The other parameters for the effective collision frequency
$\bar{\nu}_{ei}(v)$ are $\nu_{\rm ei, \max}/\nu_{\rm ei, th} = 100$ and
$v_c/\vthe = 6$. The phase space resolutions are the same as for the linear
Landau damping simulations. The physical parameters for the runs in
Fig. \ref{Fig. non-lin. Landau damp., with/without coll.} are in fact
identical to the ones considered in Fig. 9 of Ref. \cite{Brunner09}, except
that the LOKI code considers a Lorentz collision operator restricted to 2D
velocity space, while the results obtained in Ref. \cite{Brunner09} with a
collisional version of the semi-Lagrangian SAPRISTI code that supports scattering in
3D velocity space. The LOKI and SAPRISTI results are nonetheless strikingly
similar.

Also shown in Fig. \ref{Fig. non-lin. Landau damp., with/without coll.} is
the distribution of electrons in the resonant region of velocity space 
corresponding to these two cases. The addition of moderate collisions
acts to isotropize the distribution of trapped electrons and reduces the 
marked oscillations along $v_x$ at the high velocity edge 
of the trapping regions where there are steep velocity gradients. 
%

\section{Conclusions}\label{sec:conclusions}

Because a fully five dimensional phase space (2 space plus 3 velocity) is
beyond the computational resources for continuum method solution to the Vlasov
system of equations, we constructed LOKI with a four dimensional (2 space plus
2 velocity) algorithm.  A simple pitch angle scattering operator in
two-dimensional velocity space has been implemented in the LOKI code using a
conservative, finite-difference discretization scheme.  Despite the fact that
the time integration scheme in LOKI is explicit (4th order Runge-Kutta), the
CFL-type limit on the time step remains 'affordable', even at the higher end
of the relevant range of collision frequencies. A practical, approximate,
analytical relation (see Eq. (\ref{time limit, e-i collisions, explicit R-K}))
for estimating the CFL time limit related to the collision
operator was derived and validated in Appendix \ref{sec:discretization}. The implementation
has been successfully tested in the case of collisional effects on linear
Landau damping, where results obtained with an alternative numerical approach
were available for comparison. Results of non-linear Landau damping were
obtained as well. An improved collision operator,
corresponding to test particle collisions off a Maxwellian distribution, is
under development.  Although still restricted to two-dimensional velocity
space, such an operator would account for thermalization effects, currently
missing with the simple pitch angle scattering operator. 

Landau damping and electron plasma waves constitute both fundamentally
interesting plasma physics processes and the least complicated systems to
study within the context of the Vlasov-Poisson system of equations. Here, we
added the simplest collisional operator, electron-ion, pitch-angle scattering
and examined the effects on EPWs of Landau and collisional damping. Despite
its simplicity and its importance to basic plasma physics, Landau damping in
the presence of electron scattering from the ions has received little
attention. There have been several analytical theory papers motivated by early
work on plasma waves echoes\cite{Gould1967, Malmberg1968} and the effects of
collisions on echoes\cite{Baker1968, Su1968}. Recent semi-analytic work on EPW
damping with collisions\cite{Brantov12} followed the approach in
Sec. \ref{Legendre Polynomial decomposition}, included self-collisions as
well, but employed a time Fourier transform rather than direct numerical
integration as done here. 

The dependence of the collisional damping on $k \lambda_{De}$ that we found in
Fig. \ref{Collisional EPW damping} relies on the assumption that collisions do
not 'disrupt' Landau damping as we find the collisional damping merely by
subtracting the collisionless Landau rate from the total damping.  In
Sec. \ref{interpretation}, we used the linearized equations expanded in a
Fourier series obtained in Sec. \ref{2D Fourier decomposition} but modified to
exclude Landau damping in one limit and to exclude collisional damping in
another limit. We examined the case $k \lambda_{De} = 0.425$ with a strong
Landau damping because of its low phase velocity, $\vph = 3.1 \vthe$ where
electron-ion collision effects on the wave-particle resonance should be
strongest. As Fig. \ref{limits:EPW damping} shows, we found that these
equations show no decrease in Landau damping as $\nuth$ increases, in fact, a
slight increase. The sum of the collisional damping and the Landau damping
from this modified set of equations shows a total damping similar to but
slightly bigger than the LOKI results and the results from the full linearized
set. The full set and the LOKI results are in excellent agreement as shown in
Fig. \ref{Fig. Coll. effects on EPW freq all ks}c. Thus, the separation of
processes was not completely successful.

An important and unexpected outcome of our simulations and analysis is that the 
collisional damping of EPWs is less than obtained from a linearized set of fluid
equations where the electron-ion momentum exchange term is the source of the damping.
The collision term in momentum moment of the Fokker-Planck equation, Eq. (\ref{fluid eqns2}), 
includes a collision rate, $\nu_e^D$, that is strongly weighted by low velocities which
leads in turn to a EPW collisional damping rate, $\nu_{coll} \simeq \nu_e^D /2$
whereas the correct kinetic treatment in Appendix \ref{sec:diffusivedamping} 
shows that, for EPWs, the effective rate should
be weighted by much higher but still bulk velocity electrons. 

The LOKI simulations also revealed the presence of entropy modes, that is,
non-oscillatory, weakly-damped modes. The nature of these modes is sensitive to the absence of
collisional thermalization effects. Thus, in this manuscript, we emphasize only the need
to account for their presence when extracting EPW damping rates. 

We briefly studied the effect of collisions on nonlinear processes that rely on 
particle trapping. This subject will be examined more fully in subsequent publications 
for both EPWs and ion acoustic waves.

\section{Acknowledgments}
\label{thanksto}
We are pleased to acknowledge valuable discussions with T. Chapman and
B. I. Cohen.  This work was performed under the auspices of the
U.S. Department of Energy by Lawrence Livermore National Laboratory under
Contract DE-AC52-07NA27344 and funded by the Laboratory Research and
Development Program at LLNL under project tracking codes 12-ERD-061 and
15-ERD-038. Computing support for this work came from the Lawrence Livermore
National Laboratory (LLNL) Institutional Computing Grand Challenge program.

%
%
%
\clearpage
\begin{appendix}
\section{Discretization with finite differences}
\label{sec:discretization}
Here we briefly discuss the discretization of the pitch angle collision operator 
(\ref{Lorentz coll. op. in Cartesian coords.})
using fourth-order accurate conservative finite differences. For convenience 
we reproduce the operator here
\begin{align*}
  \nonumber
  C_{ei}f
  = & \,\nu_{\rm ei, th}
  \left\{
  \frac{\partial}{\partial v_x}
  \left[
    \left(
    \frac{\vthe}{v}
    \right)^3
    v_y
    \left(
    v_x\frac{\partial f}{\partial v_y}
    -v_y\frac{\partial f}{\partial v_x}
    \right)
    \right] 
  \right. \\
  & \hspace{8.mm}
  \left.
  - \frac{\partial}{\partial v_y}
  \left[
    \left(
    \frac{\vthe}{v}
    \right)^3
    v_x
    \left(
    v_x\frac{\partial f}{\partial v_y}
    -v_y\frac{\partial f}{\partial v_x}
    \right)
    \right]
  \right\}.
\end{align*}
It will be convenient to express this operator in the generic form
\begin{equation}
  \label{functional form of Cei}
  C_{ei}f=\pvx{}\left(a\pvx{f}+b\pvy{f}\right)+\pvy{}\left(c\pvx{f}+d\pvy{f}\right)
\end{equation}
with $a$, $b$, $c$, and $d$ defined respectively by 
\begin{align*}
  a & = -\bar{\nu}_{ei}(v)\,v_y^2, \\
  b & = +\bar{\nu}_{ei}(v)\,v_xv_y,  \\
  c & = +\bar{\nu}_{ei}(v)\,v_xv_y,  \\
  d & = -\bar{\nu}_{ei}(v)\,v_x^2, 
\end{align*}
where the actual velocity dependent collision frequency $\bar{\nu}_{ei}(v)$
implemented in LOKI is given explicitly in Equation (\ref{e-i collision freq.,
  capped, V2}).

The conservative treatment of the conservative unmixed and mixed derivatives
in Eq. (\ref{functional form of Cei}) will be described below. Throughout the
discussion we will use standard notation for the divided difference operators
in the $v_x$ direction:
\begin{align*}
  \Dpx f_{i,j} & = \frac{f_{i+1,j}-f_{i,j}}{\dvx}, \\
  \Dmx f_{i,j} & = \frac{f_{i,j}-f_{i-1,j}}{\dvx}, \\
  \Dzx f_{i,j} & = \frac{f_{i+1,j}-f_{i-1,j}}{2\dvx}.
\end{align*}
Likewise for the $v_y$ direction:
\begin{align*}
  \Dpy f_{i,j} & = \frac{f_{i,j+1}-f_{i,j}}{\dvy}, \\
  \Dmy f_{i,j} & = \frac{f_{i,j}-f_{i,j-1}}{\dvy}, \\
  \Dzy f_{i,j} & = \frac{f_{i,j+1}-f_{i,j-1}}{2\dvy}.
\end{align*}

\subsection{Unmixed derivative operators}
\label{subsec:unmixed}
Consider the unmixed derivative in the $v_x$-direction, in particular, 
the term $\pvx{}(a\pvx{f})$, the
formulation in the $v_y$-direction being a straightforward extension. 
Following the discussion
in Ref.~\cite{henshaw06b}, conservative discretizations can be
obtained. Begin by defining coefficients $\beta_k$ to satisfy the identity
\begin{align}
\nonumber
  &\frac{\partial w}{\partial {v_x}}\left(v_x\pm\frac{\dvx}{2},v_y\right)=\\
  \nonumber
  &\quad \Dpmx\left[\sum_{k=0}^{m}\beta_k\dvx^{2k}(\Dpx\Dmx)^k\right]w\left(v_x,v_y\right)\\
  &\quad +\Order(\dvx^{2m+2}),
  \label{eq:dpdmCoeffs}
\end{align}
for infinitely differentiable functions $w$. For our purposes we are seeking 
fourth-order accurate discretizations and therefore need only $\beta_0$ 
and $\beta_1$. However, as discussed in Ref.~\cite{fornberg96}, the $\beta$ 
coefficients can be evaluated to arbitrary order using 
\[
  \xi=\sin(\xi)\left[\sum_{k=0}^{\infty}\beta_k(-4\sin^2(\xi))^k\right],
\]
giving $\beta_0=1$, $\beta_1=-\frac{1}{24}$, $\beta_2=\frac{3}{640}$, $\beta_3=-\frac{5}{7168}$, $\beta_4=\frac{35}{294912}$, and so forth. Now define the following operators:
\[
  \SDpmx = \Dpmx\left[\sum_{k=0}^{\infty}\beta_k\dvx^{2k}(\Dpx\Dmx)^k\right],
\] 
which can be interpreted as an infinite-order-accurate representation of the
derivative at the ``half points'' [e.g. $({v_x}+\dvx/2,v_y)$]. A conservative
and symmetric $p$th order accurate discretization of the term
$(af_{v_x})_{v_x}$ can be found by expanding $\SDpx(a_{i-1/2,j}\SDmx)$ in
powers of $\dvx$ and discarding small terms. Thus to fourth-order accuracy
\begin{align}
  \nonumber
  &\left[\frac{\partial}{\partial {v_x}}\left(a\frac{\partial}{\partial {v_x}}\right)\right]_{ij}
  = 
  \beta_0\Dpx(a_{i-1/2,j}^{(4)}\Dmx)\\
  \nonumber
  & \quad+\dvx^2\left[\beta_1\Dpx(a_{i-1/2,j}^{(2)}\Dmx\Dpx\Dmx)\right.\\
  \nonumber
  & \quad\left.+\beta_1\Dpx\Dpx\Dmx(a_{i-1/2,j}^{(2)}\Dmx)\right]\\
  \label{unmixed derivative operator}
  &\quad +\Order(\dvx^4).
\end{align}
Note the appearance of $a_{i-1/2,j}^{(p)}$ indicating that $a$ is needed at
the half points $({v_x}+\dvx/2,v_y)$ with $p$th order accuracy. $a$ is assumed
to be known at nodes (or integer points) and appropriate order interpolation
formulae are needed to maintain the overall order or accuracy. 
Here we use
\begin{align*}
  a_{i-1/2,j}^{(4)} & = \frac{1}{16}\left[-a_{i-2,j}+9a_{i-1,j}+9a_{i,j}-a_{i+1,j}\right],\\
  a_{i-1/2,j}^{(2)} & = \frac{1}{2}\left[a_{i-1,j}+a_{i,j}\right].
\end{align*}

Note that there is some ambiguity in the basic definition
of the first-order operators that serve as the basis for the eventual
discretization. The choice of considering Eq. (\ref{eq:dpdmCoeffs}), in contrast to 
\begin{align}
  \nonumber
  &\frac{\partial w}{\partial {v_x}}\left(v_x,v_y\right)=\\
  \nonumber
  &\quad\Dzx\left[\sum_{k=0}^{m} \eta_k\dvx^{2k}(\Dpx\Dmx)^k\right]w\left(v_x,v_y\right)\\
  &+\quad\Order(\dvx^{2m+2}),
  \label{eq:dz2Coeffs}
\end{align}
was made to ensure the resulting discretization uses a minimal stencil and
contains no null-space. To illustrate the meaning of these statements, 
observe that the second-order accurate constant coefficient operators
obtained using the anzatz in (\ref{eq:dpdmCoeffs}) and (\ref{eq:dz2Coeffs})
are $\Dpx\Dmx f_{i,j}=(f_{i+1,j}-2f_{i,j}+f_{i-1,j})/\dvx^2$ and
$\Dzx\Dzx=(f_{i+2,j}-2f_{i,j}+f_{i-2,j})/4\dvx^2$ respectively.  The former is
the well-known second-order accurate approximation to the second-derivative
while the latter is a wide-stencil estimate of the same that would have no
effect on a function with plus-minus oscillations ({\it i.e.} the null space
of the latter operator contains the functions oscillating at the Nyquist
limit).

\subsection{Mixed derivative operators}
\label{subsec:mixed}
The case of mixed derivatives follows a very similar path as the discussion
above in Section~\ref{subsec:unmixed}. 
We use Eq. \ref{eq:dz2Coeffs} for the infinitely differentiable functions $w$. 
The $\eta$ coefficients can be evaluated to arbitrary order using 
\[
  2\xi=\sin(2\xi)\left[\sum_{k=0}^{\infty}\eta_k(-4\sin^2(\xi))^k\right],
\]
giving $\eta_0=1$, $\eta_1=-\frac{1}{6}$, $\eta_2=\frac{1}{30}$, $\eta_3=-\frac{1}{140}$, $\eta_4=\frac{1}{630}$, and so forth. The following operators are defined:
\begin{align*}
  \SDzx &= \Dzx\left[\sum_{k=0}^{\infty}\eta_k\dvx^{2k}(\Dpx\Dmx)^k\right],\\
  \SDzy &= \Dzy\left[\sum_{k=0}^{\infty}\eta_k\dvy^{2k}(\Dpy\Dmy)^k\right],
\end{align*}
which can be interpreted as infinite-order-accurate representation of
$\frac{\partial}{\partial_{v_x}}$ and $\frac{\partial}{\partial_{v_y}}$ at
$({v_x},{v_y})$. As before in Section~\ref{subsec:unmixed}, conservative 
$p$th order accurate discretization of the term $\pvx{}(b\pvy{}f)$
can be found by expanding $\SDzx(b_{i,j}\SDzy)$ and discarding small
terms. Thus to fourth-order accuracy
\begin{align}
  \nonumber
  &\left[\frac{\partial}{\partial {v_x}}\left(b\frac{\partial}{\partial {v_y}}\right)\right]_{ij} = 
  \eta_0\Dzx(b_{i,j}\Dzy) \\
  \nonumber
  &\quad +\dvy^2\eta_1\Dzx(b_{i,j}\Dzy\Dpy\Dmy)\\
  \nonumber
  &\quad +\dvx^2\eta_1\Dzx\Dpx\Dmx(b_{i,j}\Dzy)\\
  &\quad +\Order\left((\dvx^2+\dvy^2)^2\right).
    \label{mixed derivative operator}
\end{align}

Note that unlike the case for the unmixed derivative operator,
Eq. (\ref{unmixed derivative operator}) in Section~\ref{subsec:unmixed}, where
interpolations were needed to evaluate the coefficient $a$ at half grid
points, there is no need in Eq. (\ref{mixed derivative operator}) for
interpolation of $b$ since the coefficients are evaluated directly at integer grid
point where they are assumed to be known.  One may furthermore note that unlike
the case for the unmixed derivative operator, the case of the mixed
derivative has no natural choices associated with the definition of the first
derivative operators that serve as alternate foundations for the eventual
discretization. As a result there is no ambiguity in the eventual definition
of the scheme as was the case for the unmixed derivatives in
Section~\ref{subsec:unmixed}.

\subsection{Estimate of the time step (CFL) limit}
\label{Sec. Estimate of CFL limit}

The LOKI code makes use of an explicit time integration based on a
fourth-order Runge-Kutta scheme.  With respect to the collision operator, a
CFL-like constraint is thus imposed on the integration time step $\Delta t$ to
ensure numerical stability. The CFL constraint is determined by
\begin{equation}
  \label{CFL, general}
  \Delta t \le \alpha / \lambda_{\max},
\end{equation}
where $\lambda_{\max}$ stands for the maximum eigenvalue of the discretized
form of the collision operator (\ref{Lorentz coll. op.}), and $\alpha \simeq 2.7853$ for 
the fourth order scheme considered in LOKI.

Note that in the continuous case, the eigenvalues $\lambda$ given by
Eq. (\ref{eigenvalue of C_ei, restricted}) can become arbitrarily large,
either as a result of the poloidal mode number $m \to \infty$, or $v_0 \to 0$.
To avoid the issue of the singularity of the collision frequency $\nu_{\rm
  ei}(v)$ as $v\to 0$ the frequency is capped to a maximum value $\nu_{\rm ei,
  \max} = \nu_{\rm ei}(\bar{v})$, where the velocity $\bar{v} > 0$ is usually
chosen such that $\bar{v} \ll \vthe$. In any case, one should recall that the
Lorentz electron-ion pitch angle scattering operator (Eq.~\ref{Lorentz
  coll. op.})  has been derived under the approximation of vanishing
electron/ion mass ratio and is therefore not valid for $v \lesssim \vthi$,
where $\vthi$ is the ion thermal velocity. Hence, one defines the capped
electron-ion collision frequency $\bar{\nu}_{ei}(v)$ as follows:
\begin{equation}
  \label{e-i collision freq., capped}
  \bar{\nu}_{ei}(v) 
  = \min[\nu_{\rm ei}(v), \nu_{\rm ei, \max} 
  = \nu_{\rm ei}(\bar{v})].
\end{equation}

\begin{figure}
  \centering
  \includegraphics[width=0.4\textwidth]{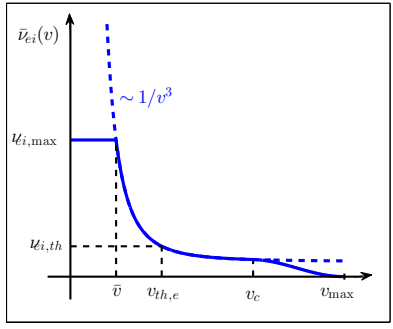}
 \caption[]{\label{Fig. e-i collision freq.} Modified electron-ion
    collision frequency as a function of velocity amplitude $v$.}
\end{figure}

In the absence of collisions, the evolution equation for the particle
distribution is given by the Vlasov equation, which is an advection
equation. For such a hyperbolic equation, the boundary conditions are of
characteristic type. Adding collisions in the form of a second order
differential (diffusion) operator represents a singular perturbation as the
system changes its nature from hyperbolic to parabolic and the domain of
dependence of a given point in the phase space-time domain goes from finite
along characteristics to infinite. Note however that the collision rate
$\nu_{\rm ei}(v)$, and therefore the magnitude of the parabolic term, decays
like $v^{-3}$ for large velocities. In order to avoid complexities associated
with the implementation of boundary conditions for a diffusion-type operator,
the collision rate is furthermore modified for all $v$ beyond some critical
velocity $v_c$ so that the collision rate at the numerical velocity boundaries
is zero. Here $\vthe \ll v_c < v_{\max}$, and $v_{\max} = \min(v_{x, \max},
v_{y, \max})$ with $v_{x, \max}$, $v_{y, \max}$ being the maximum velocities
(in absolute value) considered along $v_x$ and $v_y$ respectively.  The
collision rate actually evaluated in the code is therefore
\begin{align}
  \label{e-i collision freq., capped, V2}
  \bar{\nu}_{ei}(v) = 
  \begin{cases}
    \nu_{ei,\max}, & \mbox{if } v \le \bar{v}, \\
    \nu_{ei}(v), & \mbox{if } \bar{v}<v \le v_c, \\
    \nu_{ei}(v)\left[1-\sin^2\left(\frac{\pi}{2}\frac{v-v_c}{v_{\max}-v_c}\right)\right], & \mbox{if }  v_c < v \le v_{\max},\\
    0, & \mbox{if } v>v_{\max}.
   \end{cases}
\end{align}
With this modification, the operator is hyperbolic at the velocity boundaries
of the domain, and characteristic type boundary conditions can be further
applied, as in a collisionless case. An illustration of $\bar{\nu}_{ei}(v)$ is
provided in Fig. \ref{Fig. e-i collision freq.}. Note both the capping of
$\bar{\nu}_{ei}(v)$ at $\nu_{\rm ei, \max}$ for $v < \bar{v}$ as well as the
ramp down to zero over the interval $v_c < v < v_{\max}$. The width of the
ramp-down, $\Delta v_{\rm ramp} = v_{\max} - v_c$ is typically chosen of the
order of $\Delta v_{\rm ramp}\sim\vthe$.

Replacing the collision frequency $\nu_{\rm ei}(v)$ by
$\bar{\nu}_{ei}(v)$, 
a good estimate based on Eq. (\ref{eigenvalue of C_ei, restricted} for the 
maximum eigenvalue of the discretized
collision operator presented in Sec. \ref{subsec:mixed} is
\begin{equation}
  \label{estimate lambda_max}
  \frac{\lambda_{\max}}{\nu_{\rm ei, th}}
  \simeq 
  \pi^2\frac{\vthe^3}{\Delta v^2}
  \frac{1}{\max(\Delta v, \bar{v})},
\end{equation}
where $\Delta v$ is the velocity grid resolution. In case of unequal mesh
spacings $\Delta v_x$ and $\Delta v_y$ in the $v_x$- and $v_y$- directions
respectively, one sets $\Delta v = \min(\Delta v_x, \Delta v_y)$.  Inserting
Eq. (\ref{estimate lambda_max}) into relation (\ref{CFL, general}) leads to
the following estimate on the time constraint with respect to the collisional
dynamics:
\begin{equation}
  \label{time limit, e-i collisions, explicit R-K}
  \Delta t\,\nu_{\rm ei, th}
  \lesssim
  \frac{\alpha}{\pi^2}
  \left(
  \frac{\Delta v}{\vthe}
  \right)^2
  \frac{\max(\Delta v, \bar{v})}{\vthe}.
\end{equation}

\begin{figure}
  \centering
  \includegraphics[width=0.6\textwidth]{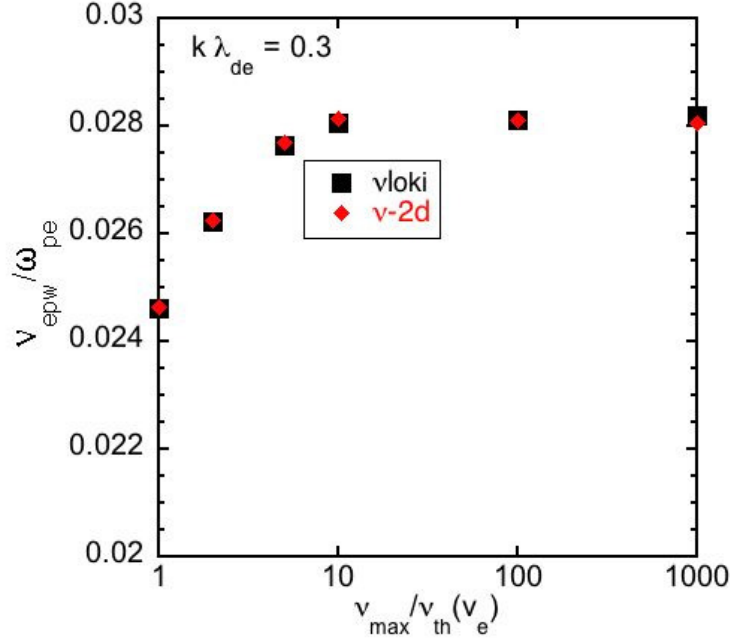}
  \caption[]{\label{convergence of damping with numax}
  The dependence of the damping rate of an EPW with $k\lambda_{De} = 0.3$ and
  $\nuth = 0.1 \omega_{pe}$ on the magnitude $\nu_{ei, \max}$ of the cap on
  $\nu_{ei}$. The results of both the LOKI simulations and the numerical
  solution to the 2D linearized set of equations are shown, both with the same
  cap.  There is no effect discernable of the cap for $\nu_{max}/\nuth >10$.
  }
\end{figure}

How the modified collision frequency profile $\bar{\nu}_{\rm ei}(v)$, and in
particular the capping at $\nu_{\rm ei, \max}$ of the maximum rate, affects
physical results is addressed in Figure \ref{convergence of damping with
numax}, where the dependence of the damping rate $\nu$ of an EPW on $\nu_{\rm
ei, \max}$ is shown. The wavenumber considered is $k\lambda_{De} = 0.3$ and
the thermal collision rate $\nuth = 1\cdot 10^{-1}\,\omega_{pe}$. According to
the results in Fig. \ref{convergence of damping with numax}, a cap set as low
as $\nu_{ei, \max} = 10\,\nu_{ei, th}$, corresponding to $\bar{v}/\vthe =
10^{-1/3} \simeq 0.464$, has no discernable effect on the EPW damping rate.

\section{Recovering the Collisional Damping of EPWs from the Linearized
  Kinetic Models}
\label{sec:diffusivedamping}

The collisional damping of EPWs, already derived in the frame of a fluid
description in Sec. \ref{collision damping:theory}, is recovered here starting
from the linearized Vlasov-Poisson systems presented in
Sec. \ref{damping:theory}. The cases of electron-ion pitch angle scattering in
three- and two-dimensional velocity space are considered in turn in the
following sections.
%
%
\subsection{Collisional Damping of EPWs in Case of 3D Velocity Scattering}
\label{diffusion 3D:theory}
Consider equations (\ref{Fokker-Planck, Legendre component l=0}) and
(\ref{Fokker-Planck, Legendre component l=1}) for the evolution of the first
two coefficients $\delta f_0(v)$ and $\delta f_1(v)$ of the fluctuating part of
the electron distribution decomposed in Legendre polynomials. For plane wave
fluctuations with frequency $\omega$ and wavenumber $k$, these two equations
read:
\begin{gather}
  \label{EWP disp.rel., collisional limit, Legendre comp. l=0}
  -i\omega\,\delta f_0
  + \frac{ikv}{3}\delta f_1
  =0,\\
  \label{EWP disp.rel., collisional limit, Legendre comp. l=1}
  -i\omega\,\delta f_1
  + ikv\,(\delta f_0 - \phi\,f_M)
  =
  - 2\,\nu_{\rm ei}(v)\,\delta f_1, 
\end{gather}
having neglected the coupling of $\delta f_1$ to $\delta f_2$ in Eq. (\ref{EWP
  disp.rel., collisional limit, Legendre comp. l=1}). This approximation is
justified in the collisonal limit, $k\lambda_{\rm ei}\ll 1$,
as the ratio of consecutive coefficients scales as $|\delta f_{l+1}/\delta
f_l| \sim k\lambda_{\rm ei}$. Equations for the higher order coefficients
$\delta f_l$, $l\le 2$, can thus be neglected. The set of equations
(\ref{EWP disp.rel., collisional limit, Legendre comp. l=0}) and (\ref{EWP
  disp.rel., collisional limit, Legendre comp. l=1}) are identical to the ones
considered in the diffusive approximation of Sec. \ref{sec:entropy}, except
for the finite inertia term $\partial\, \delta\!f_1/\partial t \sim
-i\omega\,\delta f_1$, which is clearly essential for modeling EPW dynamics.

From the set of Eqs. (\ref{EWP disp.rel., collisional limit, Legendre
  comp. l=0}) and (\ref{EWP disp.rel., collisional limit, Legendre
  comp. l=1}),  $\delta f_1$ is eliminated to obtain the following
expression for $\delta f_0$:
\[
\delta f_0(v) 
=
-\frac{(kv)^2 f_M\,\phi}{3\omega\,[\omega+2i\nu_{\rm ei}(v)]-(kv)^2},
\]
which is then inserted into the Poisson Eq. (\ref{Poisson Eq. with df0
  source}) to obtain the following dispersion relation:
\begin{equation}
  \label{3D dispersion}
  \epsilon^{3D}(k,\omega) 
  = 
  1 + \chi_e^{3D}(k,\omega) = 0,
\end{equation}
where the contribution $\chi_e^{3D}$ (corresponding to the electric
susceptibility of electrons ) to the dielectric function $\epsilon^{3D}$ reads:
\begin{equation}
  \label{3D resonant denominator}
\chi_e^{3D}(k,\omega)
= 
-\frac{1}{3}\sqrt{\frac{2}{\pi}}
\int_0^{\infty}dv\frac{v^4\exp(-v^2/2)}{\omega^2 +2i\omega \nu_{ei}(v) -k^2 v^2/3}.
\end{equation}
Assuming $\nuth \ll \omega_{pe}$, which together with the limit
$k\lambda_{\rm ei}\ll 1$ also implies $k\lambda_{De}\ll 1$, the resonant
denominator in this last relation can be Taylor expanded as follows to first
order in these small terms:
\[
\frac{1}{\omega^2 +2i\omega \nu_{ei}(v) -k^2 v^2/3}
\simeq 
\frac{1}{\omega^2} 
\left[
1 + \frac{(kv)^2}{3\omega^2} -2i\frac{\nu_{ei}(v)}{\omega}   
\right],
\]
providing, after having carried out the integral over velocity [here in
  unnormalized variables and using the relation $\nu_{ei}^{\rm brag}/\nu_{ei,th}=
  4/(3 \sqrt{2 \pi})$]:
\[
\chi_e^{3D} 
= 
-\frac{\omega_{pe}^2}{\omega^2} 
\left( 
1 + \frac{5}{3}\frac{(k\vthe)^2}{\omega^2}
-i\frac{\nu_{ei}^{brag}}{\omega} 
\right).
\]
Using this last relation, we obtain the following solution to the dispersion
relation (\ref{3D dispersion}) in the limit $k\lambda_{De} \ll 1$
and $\nuth \ll \omega_{pe}$:
\[
\omega = \omega_R - i\nu, 
\quad\text{with}\quad
\omega_R 
= 
\omega_{pe}\sqrt{1 + (5/3) (k\lambda_{De})^2}
\quad\text{and}\quad
\nu
= 
\frac{\omega_{pe}}{\omega_R}\frac{\nu_{ei}^{\rm brag}}{2}
\simeq
\frac{\nu_{ei}^{\rm brag}}{2},
\]
in full agreement with the dispersion relation (\ref{Bohm-Gross disp. rel.})
and collisional damping relation (\ref{collisional damping}) derived in
Sec. \ref{collision damping:theory}, given the relation $\gamma_e = (d+2)/d = 5/3$
 for $d=D=3$. If the thermal corrections to the dispersion are kept,
a $k \lambda_{De}$ dependence to the collisional damping is obtained. 

\begin{figure}
\begin{center}
\includegraphics[width=2.5in]{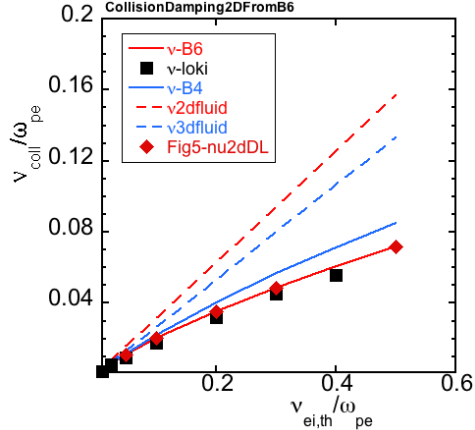}
\caption{\label{Collisional EPW damping: dispersion soln}
The collisional damping of EPWs is shown as obtained from LOKI simulations,
the linearized Fokker-Planck equations, and from fluid equations.
The solid blue and red lines are the numerical solutions for the 
collisional damping rate for Eq. (\ref{3D dispersion}) and Eq. (\ref{2D dispersion}) respectively
in the limit that $k \lambda_{ei} \to 0$. 
The LOKI results for $k \lambda_{De} =0.2 $ are shown by the black squares. The red
diamonds are the same as shown in Fig. \ref{limits:EPW damping}. They fall on the 
solid red line as they should because they are solving the same system of equations. 
The dashed red and blue lines are the damping rates from Eq. (\ref{collisional damping}) 
obtained from the 2D and 3D fluid equations respectively. Note the fluid damping rates are
larger than the simulation values and, contrary to the linearized Fokker-Planck results, larger in
2D than 3D.}
\end{center}
\end{figure}

Because $\nu_{ei}(v) \propto v^{-3} $, this Taylor expansion 
overemphasizes the contribution
of low velocities whereas the largest contribution to the real 
and imaginary part of
the integral in Eq. (\ref{3D resonant denominator}) comes 
from velocities $\sim \vthe$. In fact, the Taylor expansion to the next
order leads to a divergent integral.
A numerical solution to the dispersion Eq. (\ref{3D dispersion}) 
for the damping rate 
is shown in Fig. \ref{Collisional EPW damping: dispersion soln} 
in the limit $k \lambda_{ei} \to 0$ and is about 1/2 the rate in
Eq. (\ref{collisional damping}) derived in
Sec. \ref{collision damping:theory}.  Again, if the 
thermal corrections to the dispersion are retained,
a $k \lambda_{De}$ dependence to the collisional damping is obtained. 
A similar dependence is shown in Fig. \ref{Collisional EPW damping}. 
%
%
\subsection{Collisional Damping of EPWs in Case of 2D Velocity Scattering}
\label{diffusion 2D:theory}
We proceed as in Sec. \ref{diffusion 3D:theory}, but
start from the first two equations (\ref{Fokker-Planck, cosine component
  m=0}) and (\ref{Fokker-Planck, cosine component m=1}) for the evolution of
the first two coefficients $\delta f_{c,0}(v)$ and $\delta f_{c,1}(v)$ of the
fluctuating part of the electron distribution decomposed in polar Fourier
modes. Further making use of the Poisson equation (\ref{Poisson Eq. with dfc0
  source}), we obtain the following dispersion relation valid in the
collisional limit $k\lambda_{\rm ei}\ll 1$:
\begin{gather}
  \label{2D dispersion}
  \epsilon^{2D}(k,\omega) 
  = 
  1 + \chi_e^{2D}(k,\omega) 
  =0, \\
  \label{2D chie}
  \chi_e^{2D} 
  = 
  -\frac{1}{2} 
  \int_0^{\infty}dv
  \frac{v^3\exp(-v^2/2)}{\omega^2 +i\omega \nu_{\rm ei}-k^2 v^2/2}.
\end{gather}
After Taylor expansion of the resonant denominator, the electric
susceptibility is given by (in unnormalized variables):
\[
\chi_e^{2D} 
=
- \frac{\omega_{pe}^2}{\omega^2} 
\left( 
1 
+ 
2\frac{(k\vthe)^2}{\omega^2} 
-i\sqrt{\frac{\pi}{8}}
\frac{\nu_{ei,th}}{\omega} 
\right),
\]
Using this last relation, we obtain the following solution to the dispersion
relation (\ref{2D dispersion}) in the limit $k\lambda_{De} \ll 1$
and $\nuth \ll \omega_{pe}$:
\begin{equation}
\label{2D collisonal damping}
\omega = \omega_R - i\nu, 
\quad\text{with}\quad
\omega_R 
= 
\omega_{pe}\sqrt{1 + 2 (k\lambda_{De})^2}
\quad\text{and}\quad
\nu
= 
\frac{\omega_{pe}}{\omega_R}\sqrt{\frac{\pi}{2}}\frac{\nuth}{4}
\simeq
\sqrt{\frac{\pi}{2}}\frac{\nuth}{4},
\end{equation}
in full agreement with the dispersion relation (\ref{Bohm-Gross disp. rel.})
and collisional damping relation (\ref{collisional damping}) derived in
Sec. \ref{collision damping:theory}, where $\gamma_e = (d+2)/d = 2$ for
$d=D=2$. If the thermal corrections to the dispersion are kept,
a $k \lambda_{De}$ dependence to the collisional damping is obtained. The damping rate
in Eq. (\ref{2D collisonal damping}) is about twice larger than shown in 
Fig. \ref{Collisional EPW damping} for the reasons explained in 
Sec.\ref{diffusion 3D:theory} for the solution
of the 3D dispersion (\ref{3D dispersion}). The numerical solution to 
the 2D dispersion Eq. (\ref{2D dispersion}) and the 3D dispersion 
Eq. (\ref{3D dispersion} ) are shown in 
Fig. \ref{Collisional EPW damping: dispersion soln} along with the deduced LOKI 
collisional damping rates (also shown in Fig. \ref{Collisional EPW damping}) 
and the numerical solution to Eqs. (\ref{Fokker-Planck, cosine component m=0}) 
and (\ref{NoLD}) (also shown in Fig. \ref{limits:EPW damping}). Note the 3D damping
rate is larger than the 2D rate as expected from the kinetic results in Sec. \ref{damping:sims}.

\end{appendix}


\end{document}